\documentclass[journal]{IEEEtran}
\usepackage{graphicx}
\usepackage{amsmath}
\usepackage{amssymb}
\usepackage{array}
\newcommand{\PreserveBackslash}[1]{\let\temp=\\#1\let\\=\temp}
\newcolumntype{C}[1]{>{\PreserveBackslash\centering}p{#1}}
\newcolumntype{R}[1]{>{\PreserveBackslash\raggedleft}p{#1}}
\newcolumntype{L}[1]{>{\PreserveBackslash\raggedright}p{#1}}
\usepackage[usenames]{color}
\usepackage{colortbl,booktabs}
\usepackage{tabularx}
\usepackage{multicol}
\usepackage{booktabs}
\usepackage[Symbol]{upgreek}
\usepackage{subfigure}
\usepackage{bm}
\usepackage{cite}
\usepackage{balance}
\usepackage{mathrsfs}
\usepackage[hidelinks]{hyperref}

\usepackage{threeparttable}
\usepackage{amsmath}
\usepackage{dcolumn}
\usepackage{multirow}
\usepackage{booktabs}
\usepackage{amsfonts}

\usepackage{algorithm}
\usepackage{algpseudocode}
\usepackage{amsmath}
\usepackage{graphics}
\usepackage{epsfig}

\newcolumntype{d}[1]{D{.}{.}{#1}}

\begin{document}

\bibliographystyle{IEEEtran} 


\title{On the Power Leakage Problem in Millimeter-Wave Massive MIMO with Lens Antenna Arrays}


\author{Tian Xie,~{\it Student Member,~IEEE}, Linglong Dai,~{\it Senior Member,~IEEE}, \\Derrick Wing Kwan Ng,~{\it Senior Member,~IEEE}, and Chan-Byoung Chae,~{\it Senior Member,~IEEE}


\thanks{A part of this paper was presented at the 86-th IEEE Vehicular Technology Conference (VTC'17 Fall) \cite{xie17on}.}

\thanks{T. Xie was with the Beijing National Research Center for Information Science and Technology (BNRist), Department of Electronic Engineering, Tsinghua University, Beijing 100084, China. He is now with the China Mobile Research Institute, Beijing 100053, China (e-mail: xietian@chinamobile.com).}

\thanks{L. Dai is with the Beijing National Research Center for Information Science and Technology (BNRist), Department of Electronic Engineering, Tsinghua University, Beijing 100084, China (e-mail: daill@tsinghua.edu.cn).}

\thanks{D. W. K. Ng is with the University of New South Wales, Sydney, NSW 2052, Australia (E-mail: w.k.ng@unsw.edu.au).}

\thanks{C.-B. Chae is with the School of Integrated Technology, Yonsei University, Seoul 03722, Korea (E-mail: cbchae@yonsei.ac.kr).}

\thanks{This work was supported in part by the National Science and Technology Major Project of China under Grant 2018ZX03001004-003, and in part by the National Natural Science Foundation of China for Outstanding Young Scholars under Grant 61722109. D. W. K. Ng was supported by funding from the UNSW Digital Grid Futures Institute, UNSW, Sydney, under a cross disciplinary fund scheme and by the Australian Research Council's Discovery Project (DP190101363). C.-B. Chae was supported by IITP (No.2018-0-00170, No. 2016-0-00208).}

}

%


\maketitle
\begin{abstract}
\hspace*{0mm}
The emerging millimeter-wave (mmWave) massive multiple-input multiple-output (MIMO) with lens antenna arrays, which is also known as ``beamspace MIMO'', can effectively reduce the required number of power-hungry radio frequency (RF) chains. Therefore, it has been considered as a promising technique for the upcoming 5G communications and beyond. However, most current studies on beamspace MIMO have not taken into account the important power leakage problem in beamspace channels, which possibly leads to a significant degradation in the signal-to-noise ratio (SNR) and the system sum-rate. To this end, we propose a beam aligning precoding method to handle the power leakage problem in this paper. Firstly, a phase shifter network (PSN) structure is proposed, which enables each RF chain in beamspace MIMO to select multiple beams to collect the leakage power. Then, a rotation-based precoding algorithm is designed based on the proposed PSN structure, which aligns the channel gains of the selected beams towards the same direction for maximizing the received SNR at each user. Furthermore, we reveal some system design insights by analyzing the sum-rate and energy efficiency (EE) of the proposed beam aligning precoding method. In simulations, the proposed approach is found to achieve the near-optimal sum-rate performance compared with the ideal case of no power leakage, and obtains a higher EE than the existing schemes with either a linear or planar array.
\end{abstract}

\vspace*{0mm}

\section{Introduction}\label{S1}
\IEEEPARstart{T}he integration of millimeter wave (mmWave) communication and massive multiple-input multiple-output (MIMO), i.e., mmWave massive MIMO, has been widely considered as a promising technique in the upcoming 5G wireless communications \cite{mumtaz16mmwave,wong17key,rappaport13millimeter,xiao17millimeter}. On the one hand, the huge available bandwidths offered in mmWave bands (ranging from 30 GHz to 300 GHz) can substantially improve the throughput of wireless communications \cite{nitsche14ieee,akdeniz14millimeter}. On the other hand, the adequate array gains provided by massive MIMO \cite{rusek13} is essential to compensate the severe path loss associated with the signals in mmWave frequencies \cite{swindlehurst14millimeter,bai15coverage}. Precoding is the key to realize the considerable system throughput gain in practical mmWave massive MIMO systems \cite{Roh14millimeter}. However, the widely adopted fully digital precoding in the conventional massive MIMO (usually operating at sub-6 GHz) requires one dedicated radio frequency (RF) chain for each antenna. In particular, each RF chain includes high-resolution digital-to-analog converters \cite{gao15mmwave,heath16an}, mixers, etc. Such architecture leads to a prohibitively high energy consumption and hardware cost in mmWave massive MIMO systems for a huge number of antennas (e.g., 512 or 1024 antenna elements in \cite{heath16an,han2015large}), as the RF chains working on mmWave frequency are generally more power-hungry and costly \cite{heath16an}.

The recently proposed concept of ``beamspace MIMO'' has been considered as an effective approach to significantly reduce the number of required RF chains in mmWave massive MIMO \cite{brady13beamspace}. In beamspace MIMO systems, the lens antenna array is exploited to focus the energy of each channel path on a certain antenna element \cite{gao16low}. As a result, the traditional spatial MIMO channel can be transformed into the ``beamspace channel'', where each equivalent channel element corresponds to the gain of generating beams towards a certain direction \cite{gao16low}. One of the distinguishing properties in beamspace channels is the sparse structure thanks to the limited scattering in mmWave propagations \cite{brady13beamspace}. Therefore, only a few channel elements need to be selected for collections of most beamspace channel power, thus obviously reducing the effective dimension of the system and the number of required RF chains.

Lately, beamspace MIMO was investigated in the point-to-point MIMO systems \cite{brady13beamspace} and then extended to the multiuser scenarios \cite{sayeed13beamspace}. In \cite{zeng14Electromagnetic}, significant performance gains over conventional systems were observed in an electromagnetic lens-based multi-user uplink beamspace MIMO system. In \cite{zeng16millimeter}, the authors proposed a path division multiplexing (PDM) paradigm based on the beamspace MIMO, of which the key idea was to transmit different data streams over different paths. In \cite{zeng16multi}, the PDM paradigm intended for single-user systems was further generalized to a path division multiple access (PDMA) for multi-user scenarios. In \cite{amadori14low,amadori15low}, beam selection schemes based on different criteria were studied to improve the spectral efficiency in beamspace MIMO systems. When multiple users coincidentally shared the similar angles-of-departure (AoDs), a beam selection method was investigated in \cite{gao16near-optimal} which took into account the possible inter-user interference. The channel estimation for beamspace MIMO was studied in \cite{gao17reliable}, where a compressive sensing based method was proposed. In \cite{kwon16RF}, the authors first proposed a multi-variance codebook quantization (MVCQ) method for the limited feedback in RF lens-embedded massive MIMO systems, and then provided insights on the fabrication issues for RF lens based on an example operating on 77 GHz. Moreover, the feasibility of RF lens-embedded massive MIMO was further discussed in the continuing work \cite{cho18RF}. Two prototypes for static and mobile usage were presented and analyzed, where obvious performance gains can be observed compared with the systems without utilizing RF lens.

Despite the fruitful research in the literature, the important power leakage problem in beamspace channels is not considered in most of the current studies on beamspace MIMO, e.g., \cite{zeng14Electromagnetic,zeng16millimeter,zeng16multi,amadori14low,amadori15low,gao16near-optimal}. As the number of elements in lens antenna arrays is finite, it is impossible to always perfectly sample the randomly distributed AoDs of paths, which is illustrated in Fig. 1. Therefore, the power of some paths will inevitably disperse onto a range of antenna elements, i.e., the power leakage happens \cite{gao16low}. Conventionally, only one beam is selected for each channel path in the existing precoding approaches for beamspace MIMO systems \cite{amadori14low,amadori15low}. Hence, only a small proportion of the channel path power can be collected for information decoding, thus incurring an significant signal-to-noise ratio (SNR) and system sum-rate loss. To this end, one solution is to select multiple beams with multiple RF chains to collect sufficient channel power \cite{sayeed13beamspace,gao16near-optimal}. Although this multi-beam solution can alleviate the power leakage problem, it requires substantially more RF chains, which increases the power consumption and implementation cost of the system.

\begin{figure}[tp]
\begin{center}
\vspace*{0mm}\includegraphics[width = 1.0\linewidth]{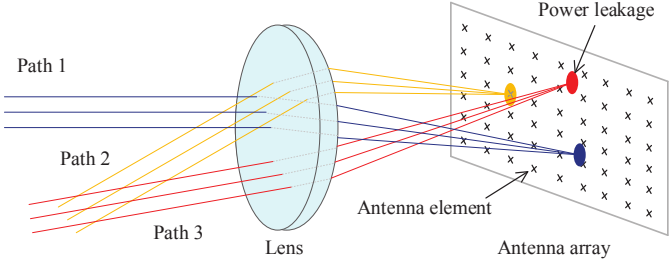}
\end{center}
\vspace*{-5mm}\caption{The path power focusing functionality of the lens and the power leakage problem in beamspace channels. }
\vspace*{0mm}
\end{figure}

In this paper, a beam aligning (BA) precoding is proposed to handle the power leakage problems\footnote{Simulation codes are provided to reproduce the results presented in this paper: \url{http://oa.ee.tsinghua.edu.cn/dailinglong/publications/publications.html}.}. The contributions of this paper can be summarized as follows:
\begin{itemize}
  \item {Firstly, we propose a phase shifter network (PSN) structure, where multiple beams can be selected via only one RF chain. Since the PSN is composed of low-cost phase shifters, it has a lower power consumption and hardware cost compared to the conventional structure exploiting multiple RF chains for selecting multiple beams.}
  \item {Then, we design a rotation-based precoding algorithm based on the proposed PSN structure. Specifically, the gains of the selected beams are aligned to the same direction through phase shifters for maximizing the received SNR for each user. The proposed rotation-based precoding algorithm enjoys a low computational complexity, as it only includes element-wise rotations which do not involve any matrix operations.}
  \item {Furthermore, we analyze the sum-rate and energy efficiency (EE) of the proposed beam aligning precoding. From the sum-rate analysis, we reveal some insights on how key system parameters affect the system sum-rate, and provide guidelines on practical system designs. It is found that the proposed precoding methods achieves a considerable EE gain compared to some conventional precoding methods.}
  \item{The proposed beam aligning precoding as well as the derived performance bounds are evaluated through simulations for both uniform linear arrays and uniform planar arrays. It can be observed that the proposed beam aligning precoding is able to achieve a near-optimal sum-rate performance compared with the ideal case of no power leakage. Besides, the proposed beam aligning precoding also has a higher EE than that of the existing schemes.}
\end{itemize}

The rest of this paper is organized as follows. The system model is described in Section II. In Section III, we propose the beam aligning precoding, where the PSN structure is firstly introduced and then the rotation-based precoding algorithm is proposed. In Section IV, we provide the sum rate and EE performance analysis of the proposed beam aligning precoding. In Section V, simulation results are provided. Finally, Section VI concludes this paper.

\emph{Notations}:  A scalar, a vector, a matrix, and a set are denoted by $a$, ${\bf{a}}$, ${\bf{A}}$, and ${\mathcal{A}}$, respectively. $\mathbb{C}$ denotes the set of all complex numbers, and $\mathbb{E}(\cdot)$ is the expectation operator for random variables. For a complex scalar $a$, $\mathbb{R} \left(a \right)$, $\mathbb{I} \left( a \right)$, and $\left| a \right|$ denote its real part, imaginary part, and the absolute value. For a vector ${\bf a}$, $\left[{\bf a}\right]_{i}$ and $|| {\bf a} ||$ denote its \emph{i}th element and the Euclidean norm, respectively. Besides, ${\bf a} \otimes {\bf b}$ is the Kronecker product of vectors ${\bf a}$ and ${\bf b}$. For a matrix ${\bf A}$, ${  {\bf{A}}  }^H$ denotes its conjugate transpose. ${\mathcal{A}} \bigcap {\mathcal{B}}$, ${\mathcal{A}} \bigcup {\mathcal{B}}$, and ${\mathcal{A}} / {\mathcal{B}}$ denote the intersection, union, and difference operation between ${\mathcal{A}}$ and ${\mathcal{B}}$. Finally, $\mathcal{N}({\bf{0}},{\bf{I}}_{K})$ ($\mathcal{CN}({\bf{0}},{\bf{I}}_{K})$) denotes the (complex) Gaussian distribution with expectation vector ${\bf{0}}$ and covariance matrix ${\bf{I}}_{K}$, where ${\bf{I}}_{K}$ is the $K \times K$ identity matrix.

\begin{figure*}[tp]
\begin{center}
\vspace*{0mm}\includegraphics[width = 1.0\linewidth]{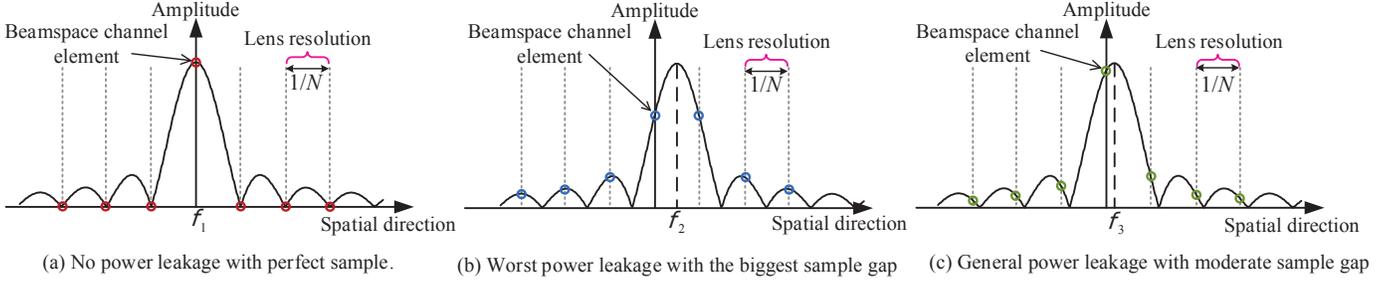}
\end{center}
\vspace*{-5mm}\caption{Illustrations of the power leakage problem in single-path case: (a) no power leakage with perfect sample; (b) worst power leakage with the biggest sample gap; (c) general power leakage with moderate sample gap.} \label{fig1}
\vspace*{0mm}
\end{figure*}

\section{System Model}
A typical mmWave beamspace MIMO system is considered, where the base station (BS) employs an $N$ element lens antenna array and $N_{\text{RF}}$ RF chains to serve $K$ single-antenna users \cite{sayeed13beamspace,gao16near-optimal}. Thus, the downlink received signal vector ${\bf{y}} \in \mathbb{C}^{K \times 1}$ at all $K$ users can be given as
\begin{equation}
{\bf{y}} = {\bf{\mathring H}}^{H}{\bf{x}} + {\bf{n}}.
\end{equation}
In equation (1), ${\bf{\mathring H}} = \left[ {\bf{\mathring h}}_1, {\bf{\mathring h}}_2, \cdots ,{\bf{\mathring h}}_{K} \right] \in \mathbb{C}^{N \times K}$ is the concatenated beamspace channel matrix with ${\bf{\mathring h}}_k \in \mathbb{C}^{N \times 1} $ presenting the individual beamspace channel vector for the \emph{k}-th user, ${\bf{x}} \in \mathbb{C}^{N \times 1}$ denotes the transmit signal vector, and ${\bf{n}} \in \mathbb{C}^{K \times 1}$ is the additive white Gaussian noise (AWGN) vector following the distribution of $\mathcal{CN} ({\bf{0}}, \sigma^2{\bf{I}}_{K})$ with $\sigma^2 {\bf{I}}_{K}$ the noise covariance matrix. Further, the total transmit power is constrained according to the total power budget $P_{\text{T}}$ as $||{\bf{x}}||^2 \leq P_{\text{T}}$. Mathematically, the lens antenna array functions as a unitary $N \times N$ discrete spatial Fourier transform matrix ${\bf{U}}$ \cite{gao16low}, where the rows are $N$ orthogonal steering vectors \cite{xiao16codebook,xiao16hierarchical}, i.e.,
\begin{equation}
{\bf{U}} = \left[ {\bf{a}}(\hat \phi_{1}), {\bf{a}}(\hat \phi_{2}), \cdots, {\bf{a}}(\hat \phi_{N})     \right]^{H},
\end{equation}
where ${\bf{a}}(\phi) \in \mathbb{C}^{N \times 1}$ is the steering vector for the spatial direction $\phi$. The spatial directions spanning the entire space are normalized and shifted as $\hat \phi_{i} = \frac{1}{N}\left(  i - \frac{N+1}{2} \right), i = 1,2,\cdots, N,$ for an expression convenience \cite{sayeed13beamspace}. Therefore, ${\bf{\mathring H}}$ can be presented as
\begin{equation}\label{beamspace}
\begin{aligned}
{\bf{\mathring H}} = {\bf{UH}} &= \left[{\bf{Uh}}_{1},{\bf{Uh}}_{2},\cdots,{\bf{Uh}}_{K}  \right],
\end{aligned}
\end{equation}
where ${\bf{H}}$ is the full dimension spatial channel matrix with ${\bf{h}}_{k} \in \mathbb{C}^{N \times 1}, k \in \left\{1, ..., k \right\}$ the individual spatial channel vector for the \emph{k}-th user.

The widely-used clustered Saleh-Valenzuela (S-V) channel model is adopted to represent the mmWave spatial channel matrix ${\bf{H}}$ \cite{el13spatially,gao16energy}:
\begin{equation}\label{channel_model}
{\bf{h}}_{k} = \sqrt{\frac{N\mu_k}{ N_{\text{cl}}^{k} N^{(k,l)}_{\text{p}}}} \sum^{N_{\text{cl}}^{k}}_{l=1}\sum^{N_{\text{p}}^{(k,l)}}_{i=1}   \beta_{k,l}^{(i)}{\bf{a}}(\phi_{k,l}^{i}).
\end{equation}
In equation (\ref{channel_model}), $\mu_k$ and $N_{\text{cl}}^{k}$ are the large-scale fading factor and the cluster number for the \emph{k}-th user, respectively. $N_{\text{p}}^{(k,l)}$ is the path number within the \emph{l}-th cluster for the \emph{k}-th user. $\beta_{k,l}^{(i)}$ and $\phi_{k,l}^{i}$ are the complex gain and the AoD corresponding to the path \emph{i}-th channel path in the \emph{l}-th channel cluster for the \emph{k}-th user, respectively.
We further assume that AoDs in the \emph{l}-th cluster, i.e., $\phi_{k,l}^{i},\  \forall i$, are uniformly distributed in a range $\left[ \phi_{k,l} - \tau_{k,l}/2, \phi_{k,l} + \tau_{k,l}/2     \right]$, where $\phi_{k,l}$ and $\tau_{k,l}$ are the averaged AoD and the angular spread corresponding to this cluster \cite{el13spatially}. If typical uniform linear arrays (ULA) with $N$ elements are considered, the steering vector will be ${\bf{a}}_{\text{ULA}}(\phi) = \frac{1}{\sqrt{N}}\left [ e^{-j2\pi\phi i} \right]_{i \in \mathcal{I}(N)}$, where the antenna indices $\mathcal{I}(N)$ are $\mathcal{I}(N) = \left\{ s - \frac{N-1}{2}, s = 0,1,\cdots, N-1\right\}$ \cite{brady13beamspace}. The spatial direction can be further defined as $\phi = \frac{\lambda}{d}\sin\theta$, where $d$ is the antenna spacing, $\lambda$ is the signal wavelength, and $\theta$ is the physical direction. Throughout this paper, we consider a half-wavelength antenna spacing, i.e., $d = \frac{\lambda}{2}$ \cite{gao16energy}.

Note that other types of antenna arrays can also be considered in the above channel model (\ref{channel_model}). For instance, if the uniform planar array (UPA) with $N_1$ horizontal elements and $N_2$ vertical elements is employed, the steering vector can be expressed as ${\bf{a}}_{\text{UPA}}(\phi_{\text{az}},\phi_{\text{el}}) = {\bf{a}}_{\text{az}}(\phi_{\text{az}}) \otimes {\bf{a}}_{\text{el}}(\phi_{\text{el}})$ \cite{he17codebook}, where ${\bf{a}}_{\text{az}}(\phi_{\text{az}})= \frac{1}{\sqrt{N_1}}\left [ e^{-j2\pi \phi_{\text{az}} m} \right]_{m \in \mathcal{I}(N_1)}$ is the azimuth steering vector with $\phi_{\text{az}}$ presenting the azimuth spatial direction, and ${\bf{a}}_{\text{el}} (\phi_{\text{el}}) = \frac{1}{\sqrt{N_2}}\left [ e^{-j2\pi \phi_{\text{el}} n} \right]_{n \in \mathcal{I}(N_2)}$ is the elevation steering vector with $\phi_{\text{el}}$ being the elevation spatial direction \cite{zeng16millimeter,zeng16multi}. The total number of antenna elements in UPA satisfies $N = N_1N_2$.

From (\ref{beamspace}), we can observe that the elements in ${\bf{\mathring h}}_{k}$ correspond to the gains of the $N$ orthogonal beams with spatial directions $\hat \phi_{i}, i = 1,2, \cdots ,N$. Due to the limited scattering characteristic for mmWave propagation \cite{gao15mmwave}, ${\bf{\mathring h}}_{k}$ has a sparse structure \cite{brady13beamspace}. Therefore, we can select the dominant beams in ${\bf{\mathring h}}_{k}$ to reduce the effective dimension of the massive MIMO systems, thus effectively reducing the number of required RF chains \cite{brady13beamspace}.

\section{Beam Aligning Precoding for Beamspace MIMO Systems}
In this section, we first explain the power leakage problem in beamspace channels. Then, we present the proposed beam aligning precoding method to handle the power leakage problem. Specifically, we first design a phase shifter network (PSN), where each RF chain is able to select multiple beams to collect sufficient path power. For the proposed PSN structure where conventional precoding algorithms cannot be applied, a rotation-based precoding algorithm is proposed to align the gains of the selected beams towards the same direction for maximizing the received SNR for each user.

\subsection{Power Leakage Problem in Beamspace Channels}
As the number of elements in lens antenna arrays is finite, it is impossible to always perfectly sample the randomly distributed AoDs of paths, which is illustrated in Fig. 1. Therefore, the power leakage in beamspace channels is inevitable. Fig. 2 illustrates three examples of the power leakage phenomena in beamspace channels, i.e., no power leakage with perfect sample in Fig. 2 (a); worst power leakage with the biggest sample gap in Fig. 2 (b); general power leakage with moderate sample gap in Fig. 2 (c). Besides, the power leakage in clustered channel is usually worse than that in single-path channel, since multiple paths in the same cluster are superimposed in a narrow range $\left[ \phi_{k,l} - \tau_{k,l}/2, \phi_{k,l} + \tau_{k,l}/2     \right]$. Note that such mismatches between spatial sample points and path AoDs have been investigated in \cite{xiao18enhanced}, where the authors proposed a two-stage channel estimation approach through a ``virtual path'' approximation of the original channel. However, beam selection in ``virtual channels'' is a step in the channel estimation algorithm, which does not require hardware devices to implement \cite{xiao18enhanced}. Therefore, selecting more beams is a natural choice since the additional computation complexity is almost negligible. In contrast, selecting beams in beamspace channels requires hardware devices such as RF chains, phase shifters, and switches to implement, leading to a complicated trade-off between the sum-rate performance and implementation costs, which will be addressed in this paper.

To quantitatively understand the power leakage problem, we provide the following computations. In a simplified single-user single-path scenario, the ratio between the leaked power and the total path power (single beam selection is assumed here) in the worst power leakage case shown in Fig. 2 (b) is
\begin{equation}\label{ratio}
\begin{aligned}
\eta =  1 - \frac{ \left| [{\bf{\mathring h}}]_{\text max} \right|^2 }{ \sum_{i = 1}^{N} \left| [{\bf{\mathring h}}]_{i} \right|^2 } = 1 - \frac{ \mathop{\max} \limits_{i} \left( {\bf{a}}^H(\hat \phi_{i}) {\bf{a}}(\phi_{\text{p}})  \right)}{ \sum_{i = 1}^{N} {\bf{a}}^H(\hat \phi_i){\bf{a}}(\phi_{\text{p}}) },
\end{aligned}
\end{equation}
where $[{\bf{\mathring h}}]_{\text{max}}$ denotes the selected beamspace channel element with the highest power (or equivalently the gain for the selcted beam), and $\phi_{\text{p}}$ is the AoD for the real channel path. For ULA, substituting ${\bf{a}}_{\text{ULA}}$ into (\ref{ratio}) yields
\begin{equation}
\begin{aligned}
\eta_{\text{ULA}} =   1-  \frac{ \mathop{\max} \limits_{\mathcal{X}_{i}} \frac{\sin^2\left( N\pi\mathcal{X}_{\text{i}}  \right)}{N^2\sin^2\left( \pi\mathcal{X}_{\text{i}}  \right)}  }{ 2\sum_{i = 1}^{N/2} \frac{\sin^2\left( N\pi\mathcal{X}_{i}  \right)}{N^2\sin^2\left( \pi\mathcal{X}_{i}  \right)}  },
\end{aligned}
\end{equation}
where ${\mathcal{X}}_{i} = \hat\phi_{i} - \phi_{\text{p}}$. In the worst power leakage case shown in Fig. 2 (b), $\mathcal{X}_{i}$ equals $(2i-1)/2N$. Thus, we have
\begin{equation}
\eta_{\text{ULA}} =  1 -\frac{1}{ 2\sum_{i = 1}^{N/2} \frac{\sin^2\left( \pi/2N  \right)}{\sin^2\left( (2i-1)\pi/2N  \right)}  }.
\end{equation}
For the UPA case, substituting ${\bf{a}}_{\text UPA}$ into (\ref{ratio}) yields
\begin{equation}
\begin{aligned}
\eta_{\text{UPA}} = 1-  \frac{ \mathop{\max} \limits_{\mathcal{X}_{\text{az},i},\mathcal{X}_{\text{el},i}} \frac{\sin^2\left( N_1\pi\mathcal{X}_{\text{az},i}  \right)}{N_1^2\sin^2\left( \pi\mathcal{X}_{\text{az},i}  \right)}  \frac{\sin^2\left( N_2\pi\mathcal{X}_{\text{el},i}  \right)}{N_2^2\sin^2\left( \pi\mathcal{X}_{\text{el},i}  \right)} }{ 4\sum \limits_{i = 1}^{N_1/2} \frac{\sin^2\left( N_1\pi\mathcal{X}_{\text{az},i}  \right)}{N_1^2\sin^2\left( \pi\mathcal{X}_{\text{az},i}  \right)} \cdot \sum \limits_{j = 1}^{N_2/2} \frac{\sin^2\left( N_2\pi\mathcal{X}_{\text{el},j}  \right)}{N_2^2\sin^2\left( \pi\mathcal{X}_{\text{el},j}  \right)} },
\end{aligned}
\end{equation}
where ${\mathcal{X}}_{\text{az},i} = \hat\phi_{\text{az},i} - \phi_{\text{az},p}$, ${\mathcal{X}}_{\text{el},i} = \hat\phi_{\text{el},i} - \phi_{\text{el},p}$, and $\phi_{\text{az},p}$ ($\phi_{\text{el},p}$) denotes the azimuth (elevation) AoD of the path. Correspondingly, in the worst power leakage scenario for UPA, $\mathcal{X}_{\text{az},i} = (2i-1)/2N_1$ and $\mathcal{X}_{\text{el},i} = (2i-1)/2N_2$. Therefore,
\begin{equation}
\begin{aligned}
&\eta_{\text{UPA}} = 1 - \frac{1}{ 4 \sum \limits_{i = 1}^{N_1/2} \frac{\sin^2\left( \pi/2N_1  \right)}{\sin^2\left( (2i-1)\pi/2N_1  \right)} \sum \limits_{i = 1}^{N_2/2} \frac{\sin^2\left( \pi/2N_2  \right)}{\sin^2\left( (2i-1)\pi/2N_2  \right)} }.
\end{aligned}
\end{equation}

Considering an example of $N = 256$ for ULA, we have $\eta_{\text{ULA}} \approx 0.60$. When $N_1 = N_2 = 16$ for UPA, we have $\eta_{\text{UPA}} \approx 0.84$, showing that most power of the path has not been collected, resulting in a substantial loss in SNR and the system sum-rate. Therefore, we will propose the beam aligning precoding to handle the power leakage problem in beamspace channels. In some literatures, the terminology ``beam alignment'' refers to the procedure of aligning one beam towards the direction of a desired user \cite{zhang17codebook,he17joint}. However, our proposed beam aligning precoding is different from that ``beam alignment'' as we align the gains instead of the directions of the selected beams. Generally, the proposed beam aligning precoding consists of two parts, i.e., the PSN structure and the rotation-based precoding algorithms, which will be discussed in following Section III-B and Section III-C, respectively.

\subsection{Proposed Phase Shifter Network Structure}
\begin{figure}[tp]
\begin{center}
\vspace*{0mm}\includegraphics[width = 1.0\linewidth]{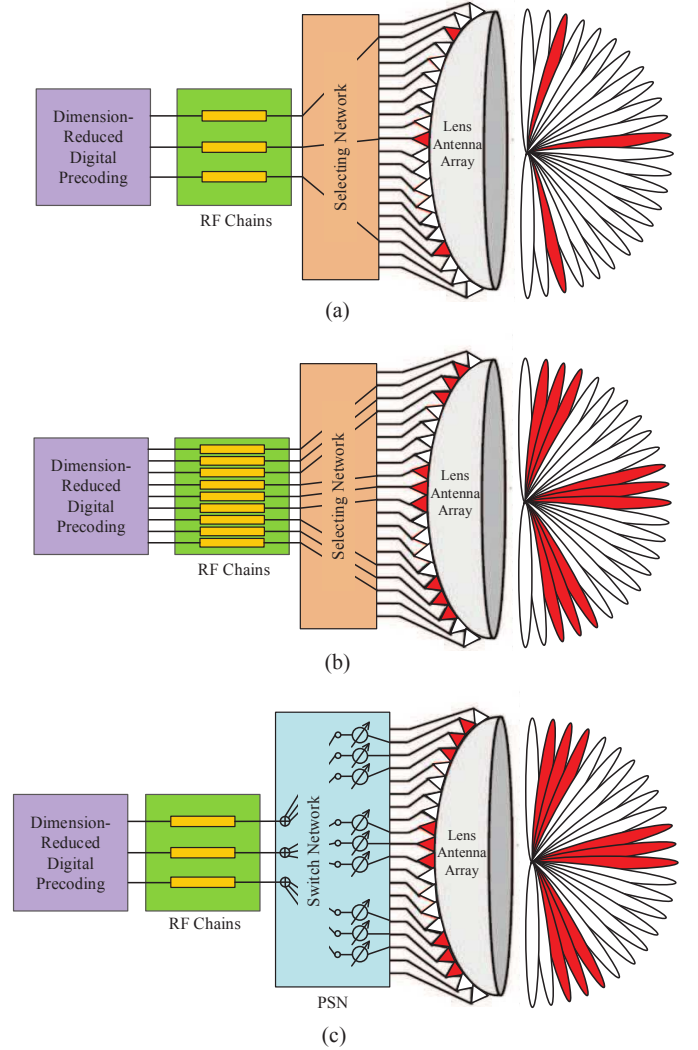}
\end{center}
\vspace*{-2mm}\caption{Precoding structure comparisons: (a) Single-beam structure; (b) MBMRF structure; (c) Proposed PSN structure.} \label{fig2}
\vspace*{0mm}
\end{figure}

In the conventional single beam (SB) precoding illustrated in Fig. 3 (a), the power leakage problem is usually omitted and only one beam is selected for each user via one RF chain \cite{amadori14low,amadori15low}, of which the power consumption model is
\begin{equation}\label{power_model_srf}
P_{\text{SB}} = P_{\text{T}} + P_{\text{BB}} + P_{\text{RF}}K + KP_{\text{SW}},
\end{equation}
where $P_{\text{BB}}$, $P_{\text{RF}}$, and $P_{\text{SW}}$ denote the power consumptions of baseband signal processing, one RF chain, and one switch, respectively. Due to the limited portion of collected path power, single beam precoding usually suffers from a severe SNR loss in beamspace channels with power leakage.

Utilizing multiple beams to collect enough path power is one solution to the power leakage problem, whose implementation structure is shown in Fig. 3 (b). Since multiple RF chains are adopted to select multiple beams \cite{sayeed13beamspace,gao16near-optimal} in such structure, we can model the power consumption of this multiple beam via multiple RF (MBMRF) structure as
\begin{equation}\label{power_model_fd}
P_{\text{MBMRF}} = P_{\text{T}} + P_{\text{BB}} + P_{\text{RF}}B_{\text{T}} + P_{\text{SW}}B_{\text{T}},
\end{equation}
where $B_{\text{T}} = \sum_{k}B_{k}$ is the total number of selected beams for all users, and $B_{k}$ represents the number of selected beams for the \emph{k}-th user. However, as RF chains in mmWave frequency are costly and power-hungry, MBMRF precoding usually incurs an exceedingly high power consumption.

To strike a balance between the system performance and the implementation cost, we propose the PSN structure to handle the power leakage problem, as shown in Fig. 3 (c). In the proposed PSN structure, each RF chain is able select multiple beams via a switch network, and $N$ phase shifters are utilized to align the gains of selected beams. To avoid the interference among users, we restrain that one beam can only be selected by one user simultaneously. Besides, the number of selected beams for each user can also be different, depending on the individual channel conditions. We can model the power consumption for the PSN structure as
\begin{equation}\label{power_model_psn}
P_{\text{PSN}} = P_{\text{T}} + P_{\text{BB}} + P_{\text{RF}}K + P_{\text{SW}}K + P_{\text{PS}}B_{\text{T}},
\end{equation}
where $P_{\text{PS}}$ is the power consumption for a phase shifter. Note that the proposed PSN is similar to the one utilized in the well-known partial connected hybrid precoding. However, we note that connecting pattern between RF chains and phase shifters (antennas) is usually fixed in the conventional PSNs, and the number of phase shifters (antennas) connecting to each RF chain is usually equally-divided (e.g., $N/K$ for each RF chain). In contrast, in the proposed PSN, the RF-phase shifter connection pattern is highly dynamic, and the number of phase shifters connecting to each RF chain can be also very different depending on the beam selection results. Therefore, conventional hybrid precoding algorithms for partial connected phase shifter networks cannot be directly applied to the proposed PSN.

It should be pointed out that substituting switches by phase shifters will generally decrease the beam switching speed in MIMO systems with lens antenna arrays. The beam switching speed for switch based structure can achieve the time-scale of nanoseconds with the emerging GaAs PIN diode switches \cite{keysight10understanding}, while that for phase shifter based structure can only reach the time-scale of microseconds \cite{bas17a}. Hence, how to reduce the beam switching time or balance it against the system performance is an important issue for the proposed PSN structure, and we would like to leave this topic for our future work.

For the proposed PSN structure, we express the transmitted signal ${\bf{x}} \in \mathbb{C}^{N \times 1}$ as
\begin{equation}
{\bf{x}} = {\bf{P}}_{\text{RF}} {\bf{P}}_{\text{BB}} {\bf{s}}.
\end{equation}
In (15), the RF domain precoder ${\bf{P}}_{\text{RF}} = \left[ {\bf{p}}^{(1)}_{\text{RF}}, {\bf{p}}^{(2)}_{\text{RF}},\cdots,{\bf{p}}^{(N_{\text{RF}})}_{\text{RF}}  \right] \in \mathbb{C}^{N \times N_{\text{RF}}}$ consists of analog precoding vectors ${\bf{p}}^{(i)}_{\text{RF}} \in \mathbb{C}^{N \times 1}$ for each RF chain, the baseband precoder ${\bf{P}}_{\text{BB}} = \left[ {\bf{p}}^{(1)}_{\text{BB}}, {\bf{p}}^{(2)}_{\text{BB}},\cdots,{\bf{p}}^{(K)}_{\text{BB}}  \right] \in \mathbb{C}^{N_{\text{RF}} \times K}$ consists of digital precoding vectors ${\bf{p}}^{(i)}_{\text{BB}} \in \mathbb{C}^{N \times 1}$ for each user, and ${\bf{s}} \in \mathbb{C}^{K \times 1}$ denotes the source information vector. Particularly, ${\bf{p}}^{(i)}_{\text{RF}}$ should satisfy
\begin{equation}\label{constraint}
\left| [ {\bf{p}}_{\text{RF}}^{(i)} ]_{j} \right| = \left\{ {\begin{array}{*{20}{c}}
{\frac{1}{\sqrt {B_i} },}&{j \in {\mathcal{B}_i},}\\
{0,}&{\text{otherwise},}
\end{array}} \right.
\end{equation}
as ${\bf{P}}_{\text{RF}}$ is implemented via phase shifters and switches in practice, where ${\mathcal{B}_i}$ contains the indices of the \emph{i}th user's selected beams. One may notice that the system model of proposed precoding structure is similar to the well-known hybrid precoding \cite{el13spatially}. However, since the observed channel matrix in beamspace MIMO is in the angular domain, the precoder designs in beamspace MIMO should directly match the sparse structure of the channel, which is different from conventional hybrid precoding methods \cite{gao16low}. Generally speaking, it is difficult to apply the conventional precoding algorithms \cite{amadori14low,amadori15low,gao16near-optimal} in the proposed PSN structure due to the different hardware constraint in (\ref{constraint}), which motivates us to design precoding algorithms specific for the proposed structure in the next subsection.

\subsection{Rotation-based Precoding Algorithm}
The received signal can be written according to (1) and (15) as
\begin{equation}
{\bf{y}} = {\bf{\mathring H}}^H{\bf{P}}_{\text{RF}} {\bf{P}}_{\text{BB}} {\bf{s}} = {\bf{\bar H}}^H{\bf{P}}_{\text{BB}} {\bf{s}},
\end{equation}
where ${\bf{\bar H}} = {\bf{P}}_{\text{RF}}^H{\bf{\mathring H}}=\left[ {\bf{\bar h}}_1,{\bf{\bar h}}_2,\cdots,{\bf{\bar h}}_K \right]$ is the equivalent RF-user domain channel. To facilitate the design of precoding,  we assume that the channel matrix ${\bf H}$ has been accurately acquired at the BS\footnote{In practice, compressive sensing based channel estimation can be employed to guarantee this assumption with low pilot overhead \cite{gao17reliable}.}. With the independent source vector ${\mathbb{E}} ( {\bf{s}}{\bf{s}}^H ) = {\bf{I}}_{K} $, the system sum-rate is \cite{gao16energy}
\begin{equation}\label{formulation1}
R = \sum^{K}_{k = 1}\text{log}_2{\left( 1+ \frac{\left|{\bf{\bar h}}^H_k {\bf{p}}^{(k)}_{\text{BB}}\right|^2}{\sigma^2 + \sum_{i \neq k} \left|{\bf{\bar h}}^H_k {\bf{p}}^{(i)}_{\text{BB}}\right|^2 }\right)}.
\end{equation}
In this paper, we consider a per-user power constraint \cite{alkhateeb16frequency} such that
\begin{equation}\label{peruser}
||{\bf{P}}_{\text{RF}}{\bf{p}}^{(k)}_{\text{BB}}||^2 \leq \frac{P_{\text{T}}}{K},\  \forall k.
\end{equation}
Thus, the precoding design problem can be formulated as maximizing the sum-rate under the constraints on the precoders
\begin{equation}\label{formulation}
\begin{array}{l}
\begin{aligned}
\left( {\bf{P}}_{\text{RF}}^{\text{opt}},{\bf{P}}_{\text{BB}}^{\text{opt}} \right) &= \mathop {\arg \max }\limits_{{\bf{P}}_{\text{RF}},{\bf{P}}_{\text{BB}}} R,\\
&\text{s.t.} \ \ {\text{(\ref{constraint})}}, \ {\text{(\ref{peruser})}}.
\end{aligned}
\end{array}
\end{equation}
Due to the existence of inter-user interference (IUI) and the non-convexity of the constraint on ${\bf{P}}_{\text{RF}}$ \cite{el13spatially,gao16energy}, it is extremely challenging to find the globally optimal solution to (\ref{formulation}). As an alternative, a sub-optimal solution through following steps is proposed.

To simplify the problem at hand, we turn to exploit the properties of the considered system. We first focus our discussions on (\ref{formulation}) to a simpler single-cluster scenario, i.e., there is only one cluster in the mmWave channel for each user, and the generalization to the multi-cluster case will be addressed in the end of this sub-section. Note that the number of BS antennas in mmWave MIMO systems is usually quite large, which is able to generate pencil beams to provide enough spatial resolution. Therefore, we could consider the average AoDs for different users $\phi_{k,l}$ separated sufficiently from each other \cite{zeng16millimeter}, which suggests that selecting multiple beams for each user will not incur significant IUI in the RF-user domain channel ${\bf{\bar H}}$.

Consequently, the IUI term in beamspace MIMO system sum-rate is not dominant \cite{zeng16millimeter,zeng16multi}, which motivates us to primarily maximizing the effective channel gains, and then suppress the IUI. Recalling that a per-user power constraint in (\ref{peruser}) is considered and temporarily ignoring the IUI in the RF-user domain channel ${\bf{\bar H}}$, we can decouple the optimization of the whole system into a sequential optimization of each user, leading to the following problem formulation for each user:
\begin{equation}\label{formulation2}
\begin{array}{l}
\begin{aligned}
\left( {\bf{P}}_{\text{RF}}^{(k),\text{opt}},{\bf{p}}_{\text{BB}}^{(k), \text{opt}} \right) &= \mathop {\arg \max }\limits_{{\bf{p}}^{(k)}_{\text{RF}},{\bf{p}}^{(k)}_{\text{BB}}}  {\left|{\bf{\bar h}}^H_k {\bf{p}}^{(k)}_{\text{BB}}\right|^2}\\
&\text{s.t.} \ \ {\text{(\ref{constraint})}}, \ {\text{(\ref{peruser})}}.
\end{aligned}
\end{array}
\end{equation}
For any given ${\bf P_{\text RF}}$, the baseband precoder maximizing the objective function in (\ref{formulation2}) is the matched filter (MF) precoder \cite{rusek13}:
\begin{equation}\label{bb}
{\bf{p}}^{(k)}_{\text{BB}} = \alpha_k {{\bf{\bar h}}_k},
\end{equation}
where $\alpha_k$ is the power normalization factor for the \emph{k}-th user. By combining (\ref{bb}) and (\ref{formulation2}), we can formulate the RF precoder design problem for the \emph{k}-th user as
\begin{equation}\label{formulation3}
\begin{array}{l}
\begin{aligned}
 {\bf{P}}_{\text{RF}}^{(k),\text{opt}}  &= \mathop {\arg \max }\limits_{ {\bf{P}}_{\text{RF}}^{(k)} } {\left|\alpha_k {\bf{\bar h}}^H_k {\bf{\bar h}}_k \right|^2}\\
&\ \ \ \text{s.t.} \ \ {\text{(\ref{constraint})}},
\end{aligned}
\end{array}
\end{equation}
where the constraint on transmit power is removed, as it can be always satisfied through adjusting $\alpha_k$. Then, we expand the objective function in (\ref{formulation3}) as
\begin{equation}\label{objective function}
\begin{aligned}
 {\left|\alpha_k {\bf{\bar h}}^H_k {\bf{\bar h}}_k \right|^2} & =   \alpha_k^2 {\left|  \left[ {\bf{\mathring h}}_k^H {\bf{p}}^{(k)}_{\text{RF}}   \right]^2 +  \sum^{K}_{j \neq k} \left[ {\bf{\mathring h}}_j^H {\bf{p}}^{(k)}_{\text{RF}} \right]^2 \right|^2} \\
 & \overset{(a)}{ \gtrsim}   \alpha_k^2 {\left| \left[ {\bf{\mathring h}}_k^H {\bf{p}}^{(k)}_{\text{RF}}   \right]^2 \right|^2},
\end{aligned}
\end{equation}
where the IUI term is omitted in the approximation $(a)$, since it is not a dominant factor as mentioned before.

Combining (\ref{formulation3}) and (\ref{objective function}), we can find that the design of RF precoder is a joint beam selection (i.e., determine the positions of non-zero elements in ${\bf{p}}^{(k)}_{\text{RF}}$) and beam combination problem (i.e., combining the selected elements in ${\bf{\mathring h}}_k$ via ${\bf{p}}^{(k)}_{\text{RF}}$). To solve the beam selection problem, we turn to examine the beamspace channel structure in the single cluster scenario. It can be observed from Fig. \ref{fig1} (c) that the beamspace channel elements are centrally distributed in an angular region corresponding to the angular spread range $\left[ \phi_{k,l} - \tau_{k,l}/2, \phi_{k,l} + \tau_{k,l}/2     \right]$ for the cluster. Therefore, we could execute the beam selection in a greedy manner within this region: Firstly, the beam with the strongest power is selected to position the cluster. Then, the beams adjacent to the previous selected beam with relatively higher leaked power are sequentially selected.

\begin{figure}[tp]
\begin{center}
\vspace*{0mm}\includegraphics[width = 1.0\linewidth]{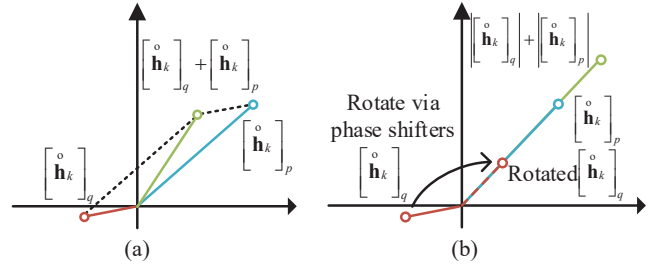}
\end{center}
\vspace*{-5mm}\caption{Illustrations of the beam gain roation procedure: (a) Combination without rotations; (b) Combination with rotations.} \label{fig3}
\vspace*{0mm}
\end{figure}

For the beam combination problem, recalling that only the phases of elements in ${\bf{p}}^{(k)}_{\text{RF}}$ are adjustable, we find the beam combination problem equivalent to rotating the selected elements in ${\bf{\mathring h}}_k$ through ${\bf{p}}^{(k)}_{\text{RF}}$ to the align their gains:
\begin{equation}\label{rf}
\frac{\left[ {\bf{p}}^{(k)}_{\text{RF}} \right]_{p}}  {\left[ {\bf{p}}^{(k)}_{\text{RF}} \right]_{q}} = \left.\left( \frac{\left[ {\bf{\mathring h}}_k \right]_{q}}{\left[ {\bf{\mathring h}}_k \right]_{p}} \right) \right/ \left| \frac{\left[ {\bf{\mathring h}}_k \right]_{q}}{\left[ {\bf{\mathring h}}_k \right]_{p}} \right|, \forall p\in {\mathcal{B}}_k,
\end{equation}
where $[ {\bf{p}}^{(k)}_{\text{RF}} ]_{q}, q \in {\mathcal{B}}_k$ is a reference element. An illustration for (\ref{rf}) is given in Fig. \ref{fig3}, where the rotated $[ {\bf{\mathring h}}_k]_{q}$ and $[ {\bf{\mathring h}}_k]_{p}$ can achieve the maximum combined value. Note that the principle of rotating beam gains towards the same direction was also investigated in [35] to address the constant-envelop precoding design. However, we would like to emphasize that the main challenge for the RF precoder design in our paper lies in the beam selection part. Once the beams are determined, the optimal combination of these beams is clear. As a result, we believe that the similarity between the beam rotation procedure and the ``geometric'' constant envelop precoding will not affect the novelty and contribution of the proposed rotation-based precoding. The overall pseudo-code for the proposed rotation-based precoding algorithm is provided in \textbf{Algorithm 1}.

\begin{algorithm}[h]
\caption{Proposed Rotation-based Precoding Algorithm}
\begin{algorithmic}[1]
\Require ${\bf{\mathring H}}$, $P_{\text{T}}$, and the beam selection threshold $\epsilon$.
\Ensure ${\bf{P}}_{\text{RF}}$, and ${\bf{P}}_{\text{BB}}$.
\State Initialize ${\mathcal{U}} = \{1,2,\cdots, N \}$, and the overall selected beam set ${\mathcal{B}} = \emptyset$;
\For {${k \leq K}$}
\State Initialize the selected beam set ${\mathcal{B}}_k = \emptyset$, the adjacent beam set ${\mathcal{A}}_k = \emptyset$, and ${\bf{p}}^{(k)}_{\text{RF}} = {\bf{0}}$ for the \emph{k}-th user;
\State $l_{\text{max}} = \text{arg max}_{m \in {\mathcal{U}}/{\mathcal{B}}} | [ {\bf{\mathring h}}_k ]_{m} |$ and ${\mathcal{B}}_k = {\mathcal{B}}_k\bigcup \{ l_{\text{max}} \}$;
\Repeat
\State Update ${\mathcal{A}}_k$ according to ${\mathcal{B}}_k$;
\State $l = \text{arg max}_{m \in {\mathcal{A}}_k } | [ {\bf{\mathring h}}_k ]_{m} |$ and ${\mathcal{B}}_k = {\mathcal{B}}_k\bigcup \{ l \}$;
\State Set $\left[ {\bf{p}}^{(k)}_{\text{RF}} \right]_{l}$ based on (\ref{rf}), where $p = l, q = l_{\text{max}}$;
\Until { $| [{\bf{\mathring h}}_k] _{l}| \leq \epsilon | [{\bf{\mathring h}}_k ]_{l_{\text{max}}} |$};
\State ${\mathcal{B}} = {\mathcal{B}} \bigcup {\mathcal{B}}_k$;
\EndFor
\State ${\bf{P}}_{\text{RF}} = [{\bf{p}}^{(1)}_{\text{RF}},...,{\bf{p}}^{(K)}_{\text{RF}} ] $;
\State ${\bf{\bar h}}^H_k = {\bf{\mathring h}}^H_k{\bf{P}}_{\text{RF}}$, $\alpha = \frac{P_{\text{T}}}{K\left\| {\bf{\bar h}}_k \right\|^2}$, ${\bf{p}}^{(k)}_{\text{BB}} = \alpha {\bf{\bar h}}_k$;
\State  ${\bf{P}}_{\text{BB}} = [{\bf{p}}^{(1)}_{\text{BB}},...,{\bf{p}}^{(K)}_{\text{BB}} ]$.
\end{algorithmic}
\end{algorithm}

In \textbf{Algorithm 1}, the RF precoder for each user is designed in a sequential manner. For each user, the algorithm first searches and selects the beam with the highest power in the selectable beam set to locate the cluster in Step 4. Then, the beams corresponding to the leaked power are selected in a greedy approach. To be specific, the neighbor beam index set ${\mathcal{A}}_k$ is updated to contain all beams adjacent to the beams in ${\mathcal{B}}_k$. The adjacent beams are defined as two different beams of which the index difference in any dimensions is at most one (as illustrated in Fig. 5). To avoid repetitive selections, we restrict that ${\mathcal{A}}_k \cap {\mathcal{B}}_k = \emptyset$. For the instance of a 2-D UPA case, we first reshape ${\bf{\mathring h}}_k$ as an $N_1 \times N_2$ matrix. Then, if ${\mathcal{B}}_k = \{ (l^{\text{az}}_{1} , l^{\text{el}}_{1}), (l^{\text{az}}_{1} , l^{\text{el}}_{1}+1)  \}$ where $l^{\text{az}}_{1}$ ($l^{\text{el}}_{1}$) is an arbitrary azimuth (elevation) index, we update ${\mathcal{A}}_k$ as
\begin{equation}
\begin{aligned}
{\mathcal{A}}_k = &\ \left\{(l^{\text{az}}_{1} , l^{\text{el}}_{1} - 1),  (l^{\text{az}}_{1} , l^{\text{el}}_{1} + 2), (l^{\text{az}}_{1} +1, l^{\text{el}}_{1}-1), \right. \\
                  &\ \ \ (l^{\text{az}}_{1} +1, l^{\text{el}}_{1}),(l^{\text{az}}_{1} +1, l^{\text{el}}_{1}+1), (l^{\text{az}}_{1} +1, l^{\text{el}}_{1}+2),\\
                  &\ \ \ (l^{\text{az}}_{1}-1 , l^{\text{el}}_{1}-1),  (l^{\text{az}}_{1}-1 , l^{\text{el}}_{1}), (l^{\text{az}}_{1} -1, l^{\text{el}}_{1}+1),\\
                  &\ \  \left. (l^{\text{az}}_{1} -1, l^{\text{el}}_{1}+2) \right\}.
\end{aligned}
\end{equation}

\begin{figure}[tp]
\begin{center}
\vspace*{0mm}\includegraphics[width = 1.0\linewidth]{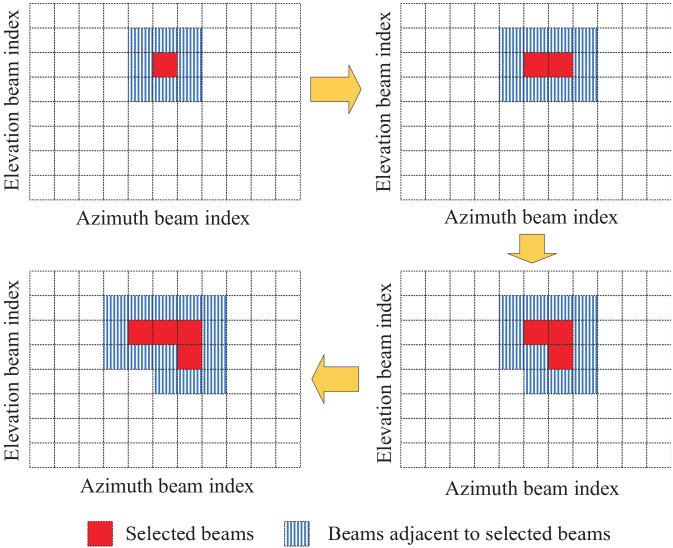}
\end{center}
\vspace*{-5mm}\caption{The greedy beam selection procedure in the UPA case.} \label{fig4}
\vspace*{0mm}
\end{figure}

After $\mathcal{A}_k$ is updated, the algorithm selects the beam with the highest power in $\mathcal{A}_k$ in step 7, and the corresponding non-zero element $[ {\bf{p}}^{(k)}_{\text{RF}} ]_{l}$ is computed according to (\ref{rf}) in step 8. The beam selection procedure (steps 5-7), which is illustrated in Fig. 5, are repeated until the power of the newly select beam $| {\bf{\mathring h}}_k |_{l}$ is smaller than a threshold $\epsilon |{\bf{\mathring h}}_k |_{l_{\text{max}}}$. When the RF precoder has been determined, the baseband precoder ${\bf P }_{\text{BB}}$ is obtained based on (\ref{bb}) and normalized in step 14. Note that we do not make any assumptions on the type of antenna arrays in this sub-section, which indicates the proposed beam-aligning precoding can be applied to any type of antenna arrays.

Finally, we analyze the computational complexity of \textbf{Algorithm 1}. To compute the RF precoder, we need to traverse the channel once, execute $B_k$ comparisons and compute $\left[ {\bf{p}}^{(k)}_{\text{RF}} \right]_{l}$ for each user, of which the complexity is $\mathcal{O}(NK + B_{\text{T}})$. For the baseband precoder, we need to compute the equivalent channel and the factor $\alpha_k$, whose the complexity is $\mathcal{O}(K^2 + B_{\text{T}})$. To sum up, the overall complexity for \textbf{Algorithm 1} is $\mathcal{O}(NK + K^2 + B_{\text{T}})$. Since $N$ is usually much larger than $K$, the main complexity comes from the beam selection, and the extra complexity brought by the computation of $P_{\text{RF}}$ and $P_{\text{BB}}$ is relatively small. However, as the computational complexity for adjacent beam searching is independent from $N$ after the beam with the highest power is found in Step 4, the overall complexity for {\bf Algorithm 1} is still kept in a low level. Basically, the complexity of proposed precoding algorithm is comparable to the single beam precoding \cite{gao16near-optimal}.

{\it Discussions for suppressing IUIs:} The above precoder design is derived by temporarily ignoring the IUI terms. However, IUI caused by shared paths/clusters among users always exists in practical systems, especially when the number of the users becomes large or $\epsilon$ is small. To address the IUI, the following solutions can be considered to modify \textbf{Algorithm~1}: 1) avoiding selecting paths that incurs obvious IUI in the RF precoder design; 2) utilizing IUI suppression baseband digital precoder, e.g., the (regularized) zero-forcing precoding \cite{rusek13}; 3) a combination of 1) and 2). For the solution of avoiding IUI through beam selection, one specific method is to replace Step~4 and Step 7 by selecting beams according to the ratio of desired signal and the interference incurred to other users $\eta$, i.e.,
\begin{equation}
l = \mathop {\arg \max }\limits_{m} \eta_m =  \mathop {\arg \max }\limits_{m} \frac{\left| \left[ {\bf \mathring h}_k \right] \right|^2_{m}}{\sigma^2 + \sum_{i \neq k}\left| \left[ {\bf \mathring h}_i \right] \right|^2_{m}},
\end{equation}
where $\sigma^2$ is the noise power. Note that the criterion for ending the algorithm in Step 8 should also be modified accordingly, e.g., ``Until $| [{\bf{\mathring h}}_k] _{l}| \leq \epsilon | [{\bf{\mathring h}}_k ]_{l_{\text{max}}} |$ or $\eta < \eta_{\text {min}}$'', where $\eta_{\text {min}}$ is a threshold to prevent incurring more interferences than the desired signals. It is worth pointing out that the greedy beam selection methods may face the problem of user unfairness when IUI occurs, since the first user to select beams enjoys a more favorable position than the follow-up users. An adaptive ending threshold for each user can be adopted to address this issue, where the users select beams with a sequentially decreased $\eta_{\text {min}}$. Note that though the above modified beam selection method could suppress the IUI, it does not necessarily lead to an optimal system-level achievable rate. If we want to achieve the optimal system-level performance, joint beam selection among all users is inevitable, which requires a prohibitively high computational complexity.

{\it Generalization to multi-cluster scenario:} The precoder designs in Section III-C are based on the assumption of a single-cluster scenario. In fact, the proposed algorithm can be generalized to multi-cluster scenario simply by changing the searching space in Step 5 from adjacent beams to all beams. However, as the number of clusters increases, the potential IUI caused by coincided paths/clusters will be severer. As a result, the straightforward generalization of \textbf{Algorithm 1} may not achieve satisfying performance in multi-cluster scenario. When the IUI is significant or even the bottleneck of the system (e.g., the paths/clusters for two users are completely coincided), handling IUI via signal processing techniques in beamspace MIMO systems is a very challenging task. In such cases, scheduling users with severe path/cluster coincidence on different orthogonal time/frequency resources could be a more effective approach. Given that designing specific user scheduling algorithms is beyond the scope of this paper, we would like to leave this topic for our future research.

\section{Performance Analysis}

In this section, we study the sum-rate and EE performance analysis on the proposed beam aligning precoding. For simplicity and without loss of generality, we consider that the ULA is adopted at the BS and omit the subscript in ${\bf{a}}(\phi)$ in this section. Note that although the derivations in this section are based on ULA, the principles can be directly generalized to arbitrary uniform arrays, such as UPA.

\subsection{Sum-rate Analysis}
To facilitate the sum-rate performance analysis, we first introduce the following \textbf{Lemma 1} to characterize the beamspace channels.

\vspace*{+1.5mm} \noindent\textbf{Lemma 1}. {\it Let the following assumptions hold:}\\
\hspace*{+2.1mm} 1) {\it The AoDs of each user are separated enough so that the IUI in the beamspace MIMO systems can be neglected \cite{zeng16multi,zeng16millimeter}.}\\
\hspace*{+2.3mm} 2) {\it The number of paths within one cluster tends to infinity, i.e., $N_{\text{p}}^{(k,l)} \rightarrow \infty,\  \forall k,l$.}\\
\hspace*{+2mm} 3) {\it The complex gains of each path ${\beta}_{k,l}^{(i)}, \forall i,k,l$ are i.i.d. random variables and follow $\mathcal{CN}(0,1)$ \cite{el13spatially}.}\\
\hspace*{+2mm} 4) {\it The AoDs of each path $\phi^{(i)}_{k,l}, \forall i$ are mutually independent and uniformly distributed in $[\phi^{\text{L}}_{k,l}, \phi^{\text{U}}_{k,l}]$. To simplify the expression, without loss of generality, we further assume that $\phi^{\text{L}}_{k,l} = \frac{1}{N}(S^0_{k,l} -\frac{N+1}{2} )$, $\phi^{\text{U}}_{k,l} = \frac{1}{N}(S^1_{k,l} -\frac{N+1}{2} )$, where $S^0_{k,l}$ and $S^1_{k,l}$ are integers and $S^1_{k,l} - S^0_{k,l} > 2 $.}\\
{\it \hspace*{+2.4mm} Then, we have}:
\begin{equation}
 [{\bf{\mathring h}}_k]_i  \sim \mathcal{CN}\left( 0, \frac{N \mu_k}{ N_{\text{cl}}^{k}} \sigma^2_{ki} \right),
\end{equation}
{\it where}
\begin{equation}
\sigma_{ki} = \frac{1}{2N( S_{k,l}^0 - S_{k,l}^1  )}\int^{\frac{i-S_{k,l}^1}{N}}_{\frac{i-S_{k,l}^0}{N}} \frac{\sin^2( N\pi\mathcal{X} )}{\sin^2( \pi\mathcal{X} )} d{\mathcal{X}}.
\end{equation}
\vspace*{+1.5mm}

\textit{Proof:} See Appendix A. \hfill\rule{6pt}{6pt}

\textbf{Lemma 1} provides a concise expression for $[{\bf{\mathring h}}_k]_i$ by considering that the number of paths within one cluster tends to infinity.
Note that the variances $\sigma_{ki}, \forall i$ are not necessary the same, since the integral range $[(i-S_{k,l}^1)/N, (i-S_{k,l}^0)/N]$ varies for different $i$. We argue that the assumption of a dense scattering environment within each cluster in {\bf Lemma 1} does not contradict the sparse structure of beamspace channel, since the number of clusters is still limited. In fact, the number of paths within each cluster will not be too small, and is usually keep at a moderate value (e.g., $10$ paths within one cluster as in \cite{el13spatially}) in the clustered channel model. Therefore, studying this asymptotical case can still provide some insights into the performance of the proposed beam aligning precoding.

However, computing $\sigma^2_{ki}$ in (28) analytically is very challenging which is generally intractable, since the integrand contains $\frac{\sin^2( N\pi\mathcal{X}^{(m)}_{k,l} )}{\sin^2( \pi\mathcal{X}^{(m)}_{k,l})}$ \cite{zhang07an}. As an alternative, we resort the use of a numerical method to obtain an approximation for $\sigma^2_{ki}$. Noting that the zeroes of the even function $f(\mathcal{X}) = \frac{\sin^2( N\pi\mathcal{X})}{\sin^2( \pi\mathcal{X} )}$ are $\frac{\Delta}{N}, \Delta = 1,2\cdots$ and $\lim_{N \rightarrow \infty} \int^{\frac{\Delta}{N}}_{\frac{\Delta + 1}{N}} f\left(\mathcal{X} \right) {\text{d}} \mathcal{X} = 0, \forall \Delta$, we obtain the following approximation:
\begin{equation}\label{app}
\begin{aligned}
\int^{\frac{1}{2}}_{\frac{-1}{2}} f\left(\mathcal{X} \right){\text{d}} \mathcal{X} \overset{}{\approx} \int^{\frac{1}{N}}_{\frac{-1}{N}} f\left(\mathcal{X} \right){\text{d}} \mathcal{X} \overset{(b)}{\approx} \frac{2f(\frac{1}{2N})}{N} = \frac{8N}{\pi^2},
\end{aligned}
\end{equation}
where the rectangle rule in numerical integration is utilized in the approximation $(b)$. In fact, we can compute that $\frac{\int^{\frac{1}{N}}_{\frac{-1}{N}} f\left(\mathcal{X} \right){\text{d}} \mathcal{X} }{\int^{\frac{1}{2}}_{-\frac{1}{2}} f\left(\mathcal{X} \right){\text{d}} \mathcal{X}}  > 90\%$ when $N = 512$, which verifies the accuracy of the approximation proposed in (\ref{app}). An important indication from (\ref{app}) is that $[{\bf{\mathring h}}_k]_i, i = S^0_{k,l} + 1, S^0_{k,l} + 2, \cdots, S^1_{k,l} - 1$ approximately have equal $\sigma^2_{ki}$. These elements, whose $\sigma^2_{ki}$ are larger than those of bound elements on $[\frac{1}{N}(S^0_{k,l} -\frac{N+1}{2} ), \frac{1}{N}(S^1_{k,l} -\frac{N+1}{2} )]$, are termed as ``central elements''. Note that the correlation between central elements has been reduced after the approximation in (\ref{app}), which makes the central elements tend to be mutually independent. By substituting (\ref{app}) into (\ref{integral}), we can obtain the variance of the central elements
\begin{equation}\label{lemma2}
\sigma^2_{ki} \approx \frac{4}{\pi^2( S_{k,l}^1 - S_{k,l}^0  )},
\end{equation}
where $i = S^0_{k,l} + 1, S^0_{k,l} + 2, \cdots, S^1_{k,l} - 1$. The larger variance for central elements implies a higher power, which indicates that the central elements are more likely to be selected by the beam selection algorithm. Note that the AoDs of paths within one cluster are assumed to be uniformly distributed in the cluster angular spread range $[\phi^{\text{L}}_{k,l}, \phi^{\text{U}}_{k,l}]$ rather than uniformly distributed in $[\-1/2, 1/2]$. Therefore, the power of beams will concentrate in the central range. Based on the central elements, we provide the following \textbf{Theorem 1}.

\vspace*{+2mm} \noindent\textbf{Theorem 1}. {\it For the proposed beam aligning precoding, given the following assumptions:}\\
\hspace*{+2mm} 1) {\it All assumptions in \textbf{Lemma 1}}.\\
\hspace*{+2mm} 2) {\it Central elements for a certain user have the same $\sigma^2_{k}$ and are mutually independent\footnote{According to Fig. 2, the elements in $[{\bf{\mathring h}}_k]$ are mutually dependent. However, we neglect such mutual dependence mainly to simplify the analysis. Besides, we would like to point out that the correlation between central elements has been reduced after the approximation in (28) due to a truncated integral range $\left[ -1/N, 1/N \right]$, which makes this assumption more reasonable.}}.\\
\hspace*{+2mm} 3) {\it $N$ and $S^1_{k,l} - S^0_{k,l}$ are large enough that all beams are selected from central elements.} \\
{\it \hspace*{+2mm} Then, an upper bound for the ergodic achievable sum rate of the system is given as}
\begin{equation}
\begin{aligned}
\mathbb{E} (R_{\text{BA}}) \leq \hat R_{\text{BA}} = \sum^{K}_{k = 1}{\log}_2{ \left( 1+  \frac{ \gamma \sigma^2_k \left(\pi B_k + 4 - \pi \right)}{ 2 \sigma^2 }   \right) },
\end{aligned}
\end{equation}
{\it where $ \gamma= \frac{P_{\text{T}}N\mu_k}{K N^k_{\text{cl}}}$ is the power normalization factor.}
\vspace*{+0.5mm}

\textit{Proof:} Since the IUI in beamspace MIMO systems does not dominant the system performance, the off-diagonal elements in ${\bf{\bar H}}$ tend to be zeroes. Thus, considering (\ref{bb}), (\ref{rf}), the power constraint, and the diagonal form of ${\bf{\bar H}}$, we rewrite (\ref{formulation1}) as
\begin{equation}
\begin{aligned}
\mathbb{E} (R_{\text{BA}}) &\overset{}{=} \mathbb{E} \left\{ \sum^{K}_{k = 1}{\log}_2{ \left( 1+  \frac{ P_{\text{T}} \left( {\bf{\bar h}}^H_k {\bf{\bar h}}_k \right)^2}{ K \sigma ^2 \left\| {\bf{P}}_{\text{RF}} {\bf{\bar h}}_k  \right\|^2}    \right)}   \right\} \\
&\overset{}{=}  \mathbb{E} \left\{ \sum^{K}_{k = 1}{\log}_2{ \left( 1+  \frac{P_{\text{T}} \left( \left[ {\bf{\bar h}}_k \right]_k \right)^4}{K \sigma ^2 \left\| \left[{\bf{\bar h}}_k \right]_k {\bf{p}}^{(k)}_{\text{RF}}   \right\|^2}    \right)}   \right\} \\
&\overset{}{=}  \mathbb{E} \left\{ \sum^{K}_{k = 1}{\log}_2{ \left( 1+  \frac{P_{\text{T}}  \left[ {\bf{\bar h}}_k \right]^4_k }{K \sigma ^2  B_k  \left[{\bf{\bar h}}_k \right]^2_k}    \right)}   \right\}.
\end{aligned}
\end{equation}
According to Jensen's inequality \cite{zi16energy}, we have
\begin{equation}\label{jensen}
\mathbb{E} (R_{\text{BA}}) \leq \hat R_{\text{BA}} =  \sum^{K}_{k = 1}{\log}_2{ \left( 1+  \frac{P_{\text{T}}}{K \sigma ^2  B_k}  \mathbb{E} \left\{ \left[{\bf{\bar h}}_k \right]^2_k \right\} \right) }.
\end{equation}
To obtain the close-form of $\mathbb{E} \left\{ \left[{\bf{\bar h}}_k \right]^2_k \right\}$, we have the following derivations:
\begin{equation}\label{expec}
\begin{aligned}
&\mathbb{E} \left( \left[{\bf{\bar h}}_k \right]^2_k \right) = \mathbb{E}\left\{ \left( \sum_{i \in \mathcal{B}_k} \left| \left[ {\bf{\mathring h}}_k \right]_i \right| \right)^2 \right\}\\
&\overset{(a)}{=} B_k \mathbb{E}\left( \left| \left[ {\bf{\mathring h}}_k \right]_i  \right|^2 \right) + B_k(B_k - 1) \mathbb{E}\left( \left| \left[ {\bf{\mathring h}}_k \right]_i  \right| \right)^2\\
&\overset{(b)}{=} {\frac{N\mu_k}{ N_{\text{cl}}^{k}}} \left[ 2\sigma^2_k B_k + \frac{\pi \sigma^2_k}{2}B_k(B_k - 1) \right],
\end{aligned}
\end{equation}
where $(a)$ follows from the assumption that $[{\bf{\mathring h}}_k]_i$ and $[{\bf{\mathring h}}_k]_j, \forall i,j$ are mutually independent, and all beams are selected from central elements, and $(b)$ comes from the raw moment for the Rayleigh distribution. Substituting (\ref{expec}) into (\ref{jensen}) yields
\begin{equation}
\hat R_{\text{BA}} =   \sum^{K}_{k = 1}{\log}_2{ \left( 1+  \frac{ P_{\text{T}} N \mu_k \sigma^2_k \left(\pi B_k + 4 - \pi \right)}{ 2 K N^k_{\text{cl}} \sigma^2 } \right) },
\end{equation}
which completes the proof. \hfill\rule{6pt}{6pt}

Following the approximation in (\ref{app}), we also obtain an approximated upper bound
\begin{equation}\label{app2}
\begin{aligned}
\hat R_{\text{BA}} \lesssim \tilde{R}_{\text{BA}} = \sum^{K}_{k = 1}{\log}_2{ \left( 1+  \frac{ \gamma (2\pi B_k + 8 - 2\pi)}{  \sigma^2 \pi^2 (S^1_{k,l} - S^0_{k,l}) } \right) }.
\end{aligned}
\end{equation}
Through the simulations in Section V, the above sum-rate upper bound (\ref{app}) is found to be tight, especially in high SNR regions, which motivates us to leverage it for discussing how will the key parameters affect the sum-rate performance as shown below.


{\it Insight 1:} According to (\ref{app2}), to efficiently achieve a higher achievable rate, we should match up the number of selected beams $B_k$ with the ``power leakage level'' $S_{k,l}^1 - S_{k,l}^0$ mentioned in \textbf{Lemma 1}. Specifically, we observe from (\ref{app2}) that for a certain user, when $S_{k,l}^1 - S_{k,l}^0$ is fixed, the achievable rate improves as the number of selected beams $B_k$ increases. However, it should be pointed out that increasing $B_k$ can not infinitely improve the achievable rate. If $B_k$ exceeds the number of the central elements, the power of the newly selected beams will attenuate rapidly, where \textbf{Theorem 1} no longer holds. In this case, increasing $B_k$ will disperse the transmit power on the beams with small power, which will degrade the achievable rate instead. Note that such rare case should be avoided in practice, as it wastes the degrees of freedom provided by the additionally utilized beams.

{\it Discussion on how does the number of antennas $N$ affect the sum-rate performance:} From (35), we can find that the sum-rate of the system goes to infinity in a linear manner as the number of antennas $N$ increases, because the channel vector for the {\it{k}}-th user is normalized to satisfy $|| {\bf h}_k || = N$ as presented in (4). However, we would like to point out that the gain of each central beam, i.e., $\sigma^2_{ki}$ in (29), will decrease as the the number of antennas $N$ becomes larger. The reason is that as the number of antennas $N$ becomes larger, the number of central beams within the fixed AoD range $[\phi^{\text{L}}_{k,l}, \phi^{\text{U}}_{k,l}]$ will also increase correspondingly. In other words, the power of each cluster will spread onto more channel elements as the number of antennas increases. To this end, we also need to increase the number of selected beams for each user, i.e., $B_k$, to match more central beams within the cluster.

{\it Comments on practical system designs:} As pointed out in \emph{Insight 1}, the number of selected beams $B_k$ should match the power leakage level $S_{k,l}^1 - S_{k,l}^0$ to achieve the full rate. However, realizing this is not a trivial task in practical systems, since acquiring the power leakage level information needs to traverse the channel. To this end, we find that the power leakage level of a cluster, which mainly depends on the geometry of the channel (similar to the AoDs of a cluster), varies much slower compared with the path gains \cite{zeng16multi}. Therefore, we only need to update the power leakage level information when the AoDs of a cluster have changed significantly, which has a lower complexity. On the other hand, if the power leakage level information can not be obtained, we can adaptively determine $B_k$ as presented in \textbf{Algorithm 1}, i.e., stopping the beam selecting procedure if the power of a newly-selected beam is smaller than a threshold.

\subsection{Energy Efficiency Analysis}
Based on the results of achievable sum-rate, we now discuss the EE performance in this part. According to the energy consumption model in Section III-B, the EE is defined as the ratio between system sum-rate and energy consumption \cite{wu17an}. To obtain some analytical results, we utilize the sum-rate upper bound (\ref{app2}) in this subsection, i.e.,
\begin{equation}\label{ee}
\text{EE} = \frac{R}{P} \approx \frac{\tilde{R}}{P},
\end{equation}
where $P$ refers to the energy consumption models in (\ref{power_model_srf}), (\ref{power_model_fd}), and (\ref{power_model_psn}). Note that in simulations, we find that the gap between the approximated EE and the exact EE (the ratio of exact sum-rate and the power consumption) is negligible in practical system operating regimes. To further analyze the system performance, we also obtain an upper bound of the sum-rate for the MBMRF precoding via the following \textbf{Proposition 1}.

\vspace*{+2mm} \noindent\textbf{Proposition 1}. {\it For the MBMRF precoding, given the assumptions in \textbf{Theorem 1}, we have the following upper bound for the ergodic achievable sum-rate:}
\begin{equation}
\begin{aligned}
\mathbb{E} (R_{\text{MBMRF}})  \leq  \hat R_{\text{MBMRF}} = \sum^{K}_{k = 1}{\log}_2{ \left( 1+  \frac{ 2 B_k \gamma \sigma^2_k }{  \sigma^2 }   \right) }.\\
\end{aligned}
\end{equation}
\vspace*{+0.5mm}

\textit{Proof:} See Appendix B. \hfill\rule{6pt}{6pt}

Based on (\ref{app}) and \textbf{Proposition 1}, we can obtain the approximated upper bound for the sum-rate achieved by MBMRF precoding
\begin{equation}\label{app3}
\begin{aligned}
\hat R_{\text{MBMRF}} \lesssim \tilde{R}_{\text{MBMRF}} = \sum^{K}_{k = 1}{\log}_2{ \left( 1+  \frac{ 8 B_k \gamma}{  \sigma^2 \pi^2 (S^1_{k,l} - S^0_{k,l}) } \right) }.
\end{aligned}
\end{equation}
For notation simplicity, $\Delta \tilde{R}_1$ denotes the sum-rate gap between different precoding schemes which is given by
\begin{equation}
\Delta \tilde{R}_1 = \tilde{R}_{\text{MBMRF}} - \tilde{R}_{\text{BA}} = \sum^{K}_{k = 1}{\log}_2 \left( { \frac{  1+  \frac{ 8 B_k \gamma}{  \sigma^2 \pi^2 (S^1_{k,l} - S^0_{k,l})}}{ 1+  \frac{ \gamma (2\pi B_k + 8 - 2\pi)}{  \sigma^2 \pi^2 (S^1_{k,l} - S^0_{k,l}) } }   } \right).
\end{equation}
In the high SNR regions, i.e., $\frac{\gamma}{\sigma^2} \gg 1$, the sum-rate gap can be approximated by
\begin{equation}\label{gap1}
\begin{aligned}
\Delta \tilde{R}_1 &\approx \sum^{K}_{k = 1}{\log}_2 \left( { \frac{ \frac{ 8 B_k \gamma}{  \sigma^2 \pi^2 (S^1_{k,l} - S^0_{k,l})}}{  \frac{ \gamma (2\pi B_k + 8 - 2\pi)}{  \sigma^2 \pi^2 (S^1_{k,l} - S^0_{k,l}) } }   } \right) \\
&= \sum^{K}_{k = 1}{\log}_2 \left( \frac{8}{2\pi + \frac{8- 2 \pi}{B_k}} \right) < \sum^{K}_{k = 1}{\log}_2 \left( \frac{4}{\pi} \right).
\end{aligned}
\end{equation}
Since ${4}/{\pi}$ is slightly larger than $1$, (\ref{gap1}) indicates that $\Delta \tilde{R}_1$ is very small in high SNR regions.

Next, we compare the power consumption, i.e., $P_{\text{MBMRF}}$ and $P_{\text{PSN}}$. Following \cite{rial16hybrid}, we model each part in $P_{\text{PSN}}$ as follows:
\begin{equation}\label{device_power_consumption}
\begin{aligned}
P_{\text{BB}} = 10P_{\text{ref}} = 200{\text{ mW}},\  &P_{\text{RF}} = 12P_{\text{ref}} = 240{\text{ mW}},\\
P_{\text{SW}} = 0.25P_{\text{ref}} = 5{\text{ mW}},\  &P_{\text{PS}} = 1.5P_{\text{ref}} = 30{\text{ mW}},
\end{aligned}
\end{equation}
where $P_{\text{ref}} = 20{\text{ mW}}$ is a reference value. For the transmit power, we adopt a typical value $P_{\text{T}} = 500{\text{ mW}} = 25P_{\text{ref}}$, where $\frac{\gamma}{\sigma^2} > 15{\text{ dB}}$. Thus, by assuming that the average number of selected beams for users is $B_k = \tilde B, \forall k$, we have
\begin{equation}
\begin{aligned}
P_{\text{MBMRF}} &= (35 + 12K\tilde B + 0.25NK \tilde B)P_{\text{ref}},\\
P_{\text{PSN}} &= (35 + 12K\tilde B + 0.25NK + 1.5K \tilde B)P_{\text{ref}}.
\end{aligned}
\end{equation}
If we consider a typical system setting where $N = 512$, $K = 8$, and $\tilde B = 5$, then we have $\frac{P_{\text{MBMRF}}}{P_{\text{PSN}}} \approx 4.64$. In other words, the proposed beam aligning precoding achieves similar sum-rate but only requires much less power consumption than the MBMRF precoding, i.e.,
\begin{equation}\label{ee_com}
\text{EE}_{\text{BA}} > \text{EE}_{\text{MBMRF}}.
\end{equation}
The main difference between the MBMRF precoding and the beam aligning precoding lies in the different ways to handle the power leakage problem. In particular, the MBMRF precoding utilizes more RF chains to select multiple beams, while the proposed beam aligning precoding adopts a phase shifter network to achieve the same goal. However, (\ref{ee_com}) reveals that utilizing the power-hungry RF chains in mmWave frequency to handle the power leakage problem in beamspace MIMO is not an energy-efficient choice. In fact, adopting analog devices with lower energy consumptions can achieve a higher EE performance.

\section{Simulation Results}
We evaluate the performance of the proposed beam aligning precoding and the corresponding theoretical analysis through simulations in this section. A typical beamspace massive MIMO system equipped with lens antenna array in mmWave bands is considered. The system bandwidth is configured as $500\text{ MHz}$, and the noise power spectral density is set to $-174  \text{ dBm/Hz}$ \cite{zi16energy}. For the mmWave MIMO channel, the clustered model introduced in (4) is employed, where the key system parameters are configured as: 1) the number of cluster for each user is assumed to be $N^k_{\text{cl}} = 1$; 2) the complex gain for each cluster $\beta^{(i)}_{k,l}, \forall i,k,l$ follow the distribution ${\mathcal{CN}}(0,1)$ \cite{el13spatially}; 3) $\phi_{k,l}, \forall k,l$ are generated based on a pre-defined set to guarantee a sufficient separation \cite{zeng16millimeter}; 4) the large-scale fading factor $\mu_k$ for the \emph{k}-th user is defined as \cite{akdeniz14millimeter}
\begin{equation}
\mu_k(\text{dB}) = 72 + 29.2\log_{10}(d) + \varrho,
\end{equation}
where $d$ denotes the distance between the BS and the user, and $\varrho \sim \mathcal{N}\left(0,8.7\right)$ is a perturbation factor. All users are assumed to be located $10{\text{ m}}$ away from the BS \cite{zi16energy}. The beam selection threshold in \textbf{Algorithm 1} is set to $\epsilon = 0.25$. Finally, we adopt the power assumption models (\ref{power_model_srf}), (\ref{power_model_fd}), and (\ref{power_model_psn}) for the corresponding scheme, while the power consumptions of devices follow (\ref{device_power_consumption}) \cite{dai19hybrid}.

\subsection{ULA Case}
We first consider that the ULA is equipped at the BS, where the BS utilizes an $N=512$-element ULA to serve $K = 8$ users simultaneously. Besides, the path angles $\phi^i_{k,l}, \forall i$ are assumed to follow a uniform distribution in $\left[ \phi_{k,l} - 5/N, \phi_{k,l} + 5/N  \right]$  \cite{gao15mmwave}.

\begin{figure}[tp]
\begin{center}
\vspace*{0mm}\includegraphics[width = 1.05\linewidth]{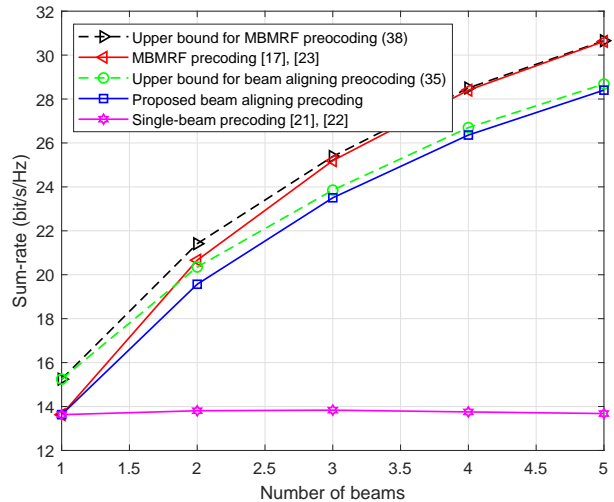}
\end{center}
\vspace*{-5mm}\caption{Sum-rate comparison versus the number of beams for the ULA case.} \label{se_beams}
\vspace*{0mm}
\end{figure}

First, we simulate the dense scattering situations within one cluster, i.e., $N^{(k,l)}_{\text{P}} = 100, \forall k,i,$ to verify the theoretical analysis derived in Section IV. To reveal the relationship between the number of selected beams for each user and the system sum-rate, we slightly modify the stop criteria of beam selection in \textbf{Algorithm 1} to ``If the number of selected beams for the \emph{k}-th user is larger than the predefined $\hat B_k$, the beam selection ends''. The sum-rate performance comparison is presented In Fig. \ref{se_beams}, where the transmit power is $P_{\text{T}} = 10 {\text{ dBm}}$ and the performance upper bounds for MBMRF precoding and proposed beam aligning precoding could be referred to (\ref{app2}) and (\ref{app3}), respectively. From Fig. \ref{se_beams}, we can observe that the derived upper bound in (\ref{app2}) is very tight, which verifies the accuracy of our analysis.

\begin{figure}[tp]
\begin{center}
\vspace*{0mm}\includegraphics[width = 1.05\linewidth]{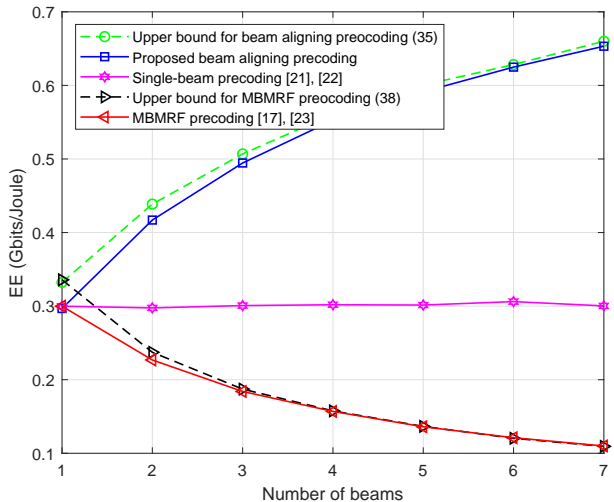}
\end{center}
\vspace*{-5mm}\caption{EE comparison versus the number of beams for the ULA case.} \label{eebeams}
\vspace*{0mm}
\end{figure}
In addition, we also provide the EE comparison result against the number of selected beams in Fig. \ref{eebeams}. The two EE upper bounds for proposed beam aligning precoding and MBMRF precoding plotted in Fig. \ref{eebeams} are obtained through substituting the derived sum-rate bounds in (\ref{app2}) and (\ref{app3}) to the EE definition in (\ref{ee}). Fig. \ref{eebeams} presents that the EE for the proposed beam aligning precoding increases as more beams are selected, since only additional phase shifters with low power consumption are utilized. In contrast, the EE for MBMRF precoding degrades as the number of selected beams gets larger, since an exceedingly high power consumption is required to drive the RF chains. Meanwhile, when more than one beam is selected, the EE for single-beam precoding is generally higher than that for MBMRF precoding, but lower than that for proposed beam-aligning precoding.

\begin{figure}[tp]
\begin{center}
\vspace*{0mm}\includegraphics[width = 1.05\linewidth]{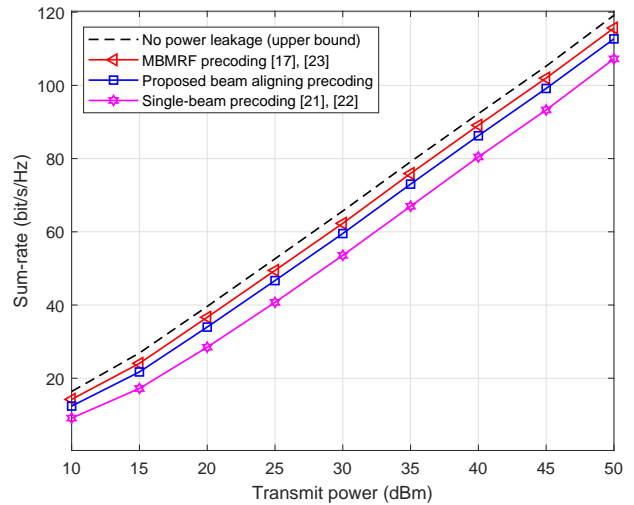}
\end{center}
\vspace*{-5mm}\caption{Sum-rate comparison versus transmit power for the ULA case.} \label{se_ula}
\vspace*{0mm}
\end{figure}

\begin{figure}[tp]
\begin{center}
\vspace*{0mm}\includegraphics[width = 1.05\linewidth]{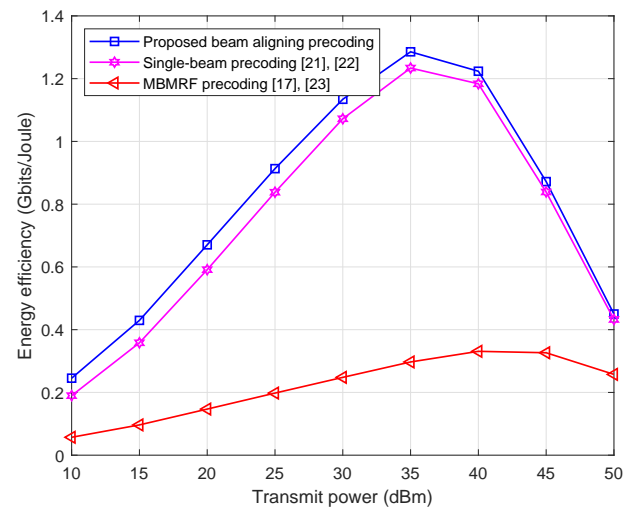}
\end{center}
\vspace*{-5mm}\caption{EE comparison versus transmit power for the ULA case.} \label{ee_ula}
\vspace*{0mm}
\end{figure}

Next, we simulate the case that the scattering within one cluster is limited, where $N^{(k,l)}_{\text{P}} = 10, \forall k,i$, is considered \cite{el13spatially}. The sum-rate performance comparison is given in Fig. \ref{se_ula}. It can be seen that the sum-rate performance of the MBMRF precoding and the proposed beam aligning precoding is higher compared to that of the single-beam precoding and approaches the ideal situation with no power leakage. In other words, selecting multiple beams is an efficient way to handle the power leakage problem. Since using multiple RF chains to select multiple beams can provide higher degrees of freedom than using a phase shifter network, it is expected that the the MBMRF precoding slightly outperforms the proposed beam aligning precoding in terms of the sum-rate.

In addition, the EE performance is also evaluated, which is given in Fig. \ref{ee_ula}. It can be found that though the MBMRF precoding can achieve a slightly higher sum-rate than other precoding methods, its EE performance severely degrades and is the worst among all considered approaches. The reason is that more RF chains in MBMRF precoding significantly increases the power consumption, which outweighs the associated sum-rate gain leading to a low EE. In contrast, the proposed beam aligning precoding achieves a higher EE, since it can achieve the near-optimal sum-rate while requiring a substantially low power consumption as presented in Section IV-B. Besides, another observation from Fig. \ref{ee_ula} is that there exists an optimal system EE operating point. Such trend can be interpreted as follows. When the transmit power is relatively small, increasing the transmit power results in a higher EE since the total power consumption is dominated by the circuit power consumption (e.g., $P_{\text{BB}}$ in (\ref{power_model_srf}), (\ref{power_model_fd}), and (\ref{power_model_psn})). However, when the transmit power is sufficiently larger than other terms in the power consumption model (e.g., $35 {\text{ dBm}}$ in Fig. \ref{ee_ula}), increasing transmit power will deteriorate EE instead. This is because the transmit power contributes to the sum-rate only in a logarithm manner, while it contributes to the total power in a linear manner.

\subsection{UPA Case}
In this subsection, we investigate the performance comparisons in the UPA case, where the BS employs a UPA with $N_1 = 32$ horizontal antenna elements and $N_2 = 16$ vertical antenna elements to serve $K = 8$ users. The total number of antennas is $N = N_1 \times N_2 = 512$. The horizontal AoDs are uniformly distributed in $\left[ \phi^{\text{az}}_{k,l} - 1/N_1, \phi^{\text{az}}_{k,l} + 1/N_1  \right]$, while the vertical AoDs are uniformly distributed in $\left[ \phi^{\text{el}}_{k,l} - 1/N_2, \phi^{\text{el}}_{k,l} + 1/N_2  \right]$. We consider a limited scattering environment where $N^{(k,l)}_{\text{P}} = 10, \forall k,i$.

\begin{figure}[tp]
\begin{center}
\vspace*{0mm}\includegraphics[width = 1.05\linewidth]{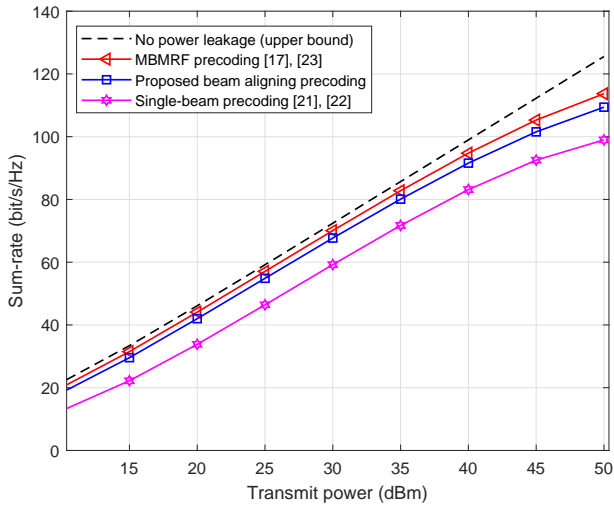}
\end{center}
\vspace*{-5mm}\caption{Sum-rate comparison versus transmit power with UPA in the limited scattering environment.} \label{se_upa}
\vspace*{0mm}
\end{figure}

\begin{figure}[tp]
\begin{center}
\vspace*{0mm}\includegraphics[width = 1.05\linewidth]{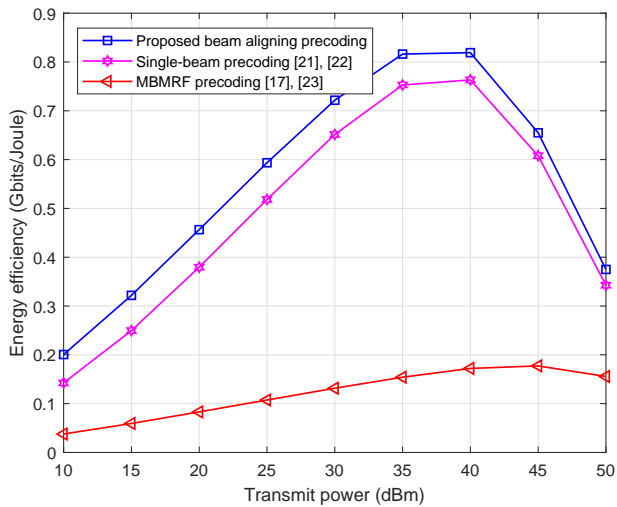}
\end{center}
\vspace*{-5mm}\caption{EE comparison versus transmit power with UPA in the limited scattering environment.} \label{ee_upa}
\vspace*{0mm}
\end{figure}

The sum-rate comparison is given in Fig. \ref{se_upa}. We can observe that the proposed beam aligning precoding is able to achieve the near-optimal sum-rate performance compared with the ideal case with no power leakage, which indicates the proposed beam aligning precoding can effectively reduce the potential power leakage. Moreover, in Fig. \ref{ee_upa}, we also provide the EE comparison in the UPA case, where the proposed beam aligning precoding is illustrated to enjoy a higher EE than the existing methods. In addition, we can find that the performance gap between single-beam precoding and the proposed beam aligning precoding in the UPA cases (Fig. \ref{se_upa}, Fig. \ref{ee_upa}) is more obvious than that in the ULA cases (Fig. \ref{se_ula}, Fig. \ref{ee_ula}), which indicates that the power leakage is more severe in channels with UPA. This is because the path power is leaked along both the vertical and horizontal dimension in channels with UPA.

\begin{figure}[tp]
\begin{center}
\vspace*{0mm}\includegraphics[width = 1.05\linewidth]{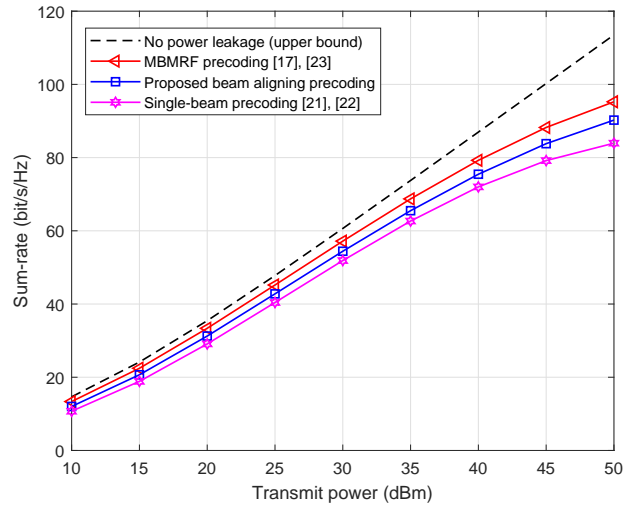}
\end{center}
\vspace*{-5mm}\caption{Sum-rate comparison versus transmit power with UPA in LoS environment.} \label{se_upa_los}
\vspace*{0mm}
\end{figure}

\begin{figure}[tp]
\begin{center}
\vspace*{0mm}\includegraphics[width = 1.05\linewidth]{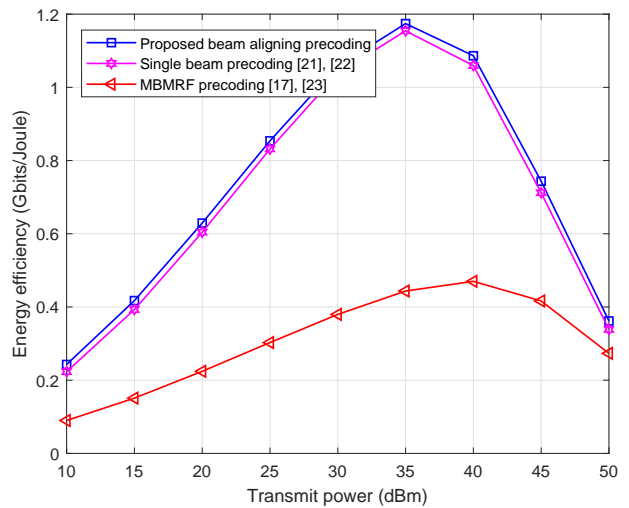}
\end{center}
\vspace*{-5mm}\caption{EE comparison versus transmit power with UPA in LoS environment.} \label{ee_upa_los}
\vspace*{0mm}
\end{figure}

Besides, we also consider the line-of-sight (LoS) environment with $N^{(k,l)}_{\text{P}} = 1, \forall k,i$, which is another typical scenario in mmWave communications \cite{brady13beamspace,gao16low,gao17reliable}. The sum-rate results are presented in Fig. \ref{se_upa_los}, from which we find that the proposed beam aligning precoding can also achieve the near-optimal performance. However, it can be observed that the performance gap between beam aligning precoding and single beam precoding becomes smaller compared with that in Fig. \ref{se_upa_los}. In fact, the power leakage problem is less severe in the LoS environment than that in the limited scattering environment. For the EE comparison illustrated in Fig. \ref{ee_upa_los}, the proposed beam aligning precoding can also achieve the highest EE performance, but the performance gain over the single beam precoding diminishes since the sum-rate advantage of the proposed beam aligning precoding over the single-beam precoding shrinks due to the alleviated power leakage in the LoS environments.

\section{Conclusions}
In this paper, we proposed a beam aligning precoding to handle the power leakage problem in mmWave massive MIMO systems with lens antenna array. The main idea of the proposed beam aligning precoding is to enable each RF chain to select multiple beams simultaneously via a phase shifter network, thus collecting sufficient path power for information decoding. Our analysis demonstrates that adopting more RF chains to select multiple beams is not an energy-efficient approach, while utilizing analog devices with lower power consumptions will achieve higher EE performance. Through simulations, we verify the ability of the proposed beam aligning precoding to efficiently handle the power leakage problem. In addition, the proposed beam aligning precoding also demonstrates a higher EE performance than conventional approaches, which is consistant with our analysis. For the future work, we will consider the power leakage problem in mmWave massive MIMO systems where users are also equipped with multiple antennas and multiple RF chains.

\section*{Appendix A \\ Proof of {\textbf{Lemma 1}}}
We only consider the contributions of paths within the same cluster and neglect the contributions from paths in other cluster since all clusters are separated from each other sufficiently far \cite{zeng16multi}. Combining (2), (3), and (4), we have
\begin{equation}\label{element}
\begin{aligned}
\left[ {\bf{\mathring h}}_k \right]_i &= \sqrt{\frac{N \mu_k}{ N_{\text{cl}}^{k} N^{(k,l)}_{\text{p}}}}\sum^{N^{(k,l)}_{\text{p}}}_m \beta^{(m)}_{k,l} {\bf{a}}^H (\frac{2i-N-1}{2N} )  {\bf{a}} ( \phi^{(m)}_{k,l} ) \\
                                      =& \sqrt{\frac{N \mu_k}{ N_{\text{cl}}^{k} N^{(k,l)}_{\text{p}}}}   \sum^{N^{(k,l)}_{\text{p}}}_m \beta^{(m)}_{k,l} \sum_{u} \frac{1}{N}e^{j2\pi ( \frac{2i-N-1}{2N} - \phi^{(m)}_{k,l} ) u} \\
                                      =& \sqrt{\frac{N \mu_k}{ N_{\text{cl}}^{k} N^{(k,l)}_{\text{p}}}}   \sum^{N^{(k,l)}_{\text{p}}}_m \beta^{(m)}_{k,l} \frac{\sin\left( N\pi\left( \frac{2i-N-1}{2N} - \phi^{(m)}_{k,l} \right)  \right)}{N\sin\left( \pi\left( \frac{2i-N-1}{2N} - \phi^{(m)}_{k,l} \right)  \right)} \\
                                      =& \sqrt{\frac{N \mu_k}{ N_{\text{cl}}^{k} N^{(k,l)}_{\text{p}}}}   \sum^{N^{(k,l)}_{\text{p}}}_m \beta^{(m)}_{k,l} \frac{\sin\left( N\pi\mathcal{X}^{(m)}_{k,l}  \right)}{N\sin\left( \pi\mathcal{X}^{(m)}_{k,l}  \right)},
\end{aligned}
\end{equation}
where $\mathcal{X}^{(m)}_{k,l} = \frac{2i-N-1}{2N} - \phi^{(m)}_{k,l}$. When $N_{\text{p}}^{(k,l)} \rightarrow \infty,\  \forall k,l$, recalling the assumption that $\beta^{(m)}_{k,l}$ are mutually independent and $\mathcal{X}^{(m)}_{k,l}$ are also mutually independent, we could apply the central limit theorem \cite{sayeed02deconstructing} to obtain $\mathbb{R} ( [{\bf{\mathring h}}_k]_i ) \sim \mathcal{N}(\zeta, \frac{N \mu_k}{ N_{\text{cl}}^{k}} \sigma^2_{ki}) ,\ \  \mathbb{I} ( [{\bf{\mathring h}}_k]_i ) \sim \mathcal{N}(\zeta,\frac{N \mu_k}{ N_{\text{cl}}^{k}} \sigma^2_{ki}),$ where $\zeta = \mathbb{E} [ ( \mathcal{R}(\beta_{k,l}^{(m)}) \frac{\sin( N\pi\mathcal{X}^{(m)}_{k,l} )}{N\sin( \pi\mathcal{X}^{(m)}_{k,l}  )} )] = 0$, and $\sigma^2_{ki} = \mathbb{E} [ ( \mathcal{R}(\beta_{k,l}^{(m)}) \frac{\sin( N\pi\mathcal{X}^{(m)}_{k,l} )}{N\sin( \pi\mathcal{X}^{(m)}_{k,l}  )} )^2]$. Next, we expand $\sigma^2_{ki}$ as
\begin{equation}\label{integral}
\begin{aligned}
\sigma^2_{ki} &= \mathbb{E} \left\{ \left( \mathcal{R}\left(\beta_{k,l}^{(m)}\right) \right)^2 \right\} \mathbb{E} \left\{ \left(\frac{\sin( N\pi\mathcal{X}^{(m)}_{k,l} )}{N\sin( \pi\mathcal{X}^{(m)}_{k,l}  )} \right)^2 \right\} \\
 &= \frac{1}{2} \mathbb{E} \left\{ \frac{\sin^2( N\pi\mathcal{X}^{(m)}_{k,l} )}{N^2\sin^2( \pi\mathcal{X}^{(m)}_{k,l}  )}  \right\}\\
 & \overset{}{=} \frac{-1}{2(\phi^{\text{U}}_{k,l} - \phi^{\text{L}}_{k,l})} \int^{\frac{2i-N-1}{2N} - \phi^{\text{U}}_{k,l}}_{\frac{2i-N-1}{2N} - \phi^{\text{L}}_{k,l}} \frac{\sin^2( N\pi\mathcal{X} )}{N^2\sin^2( \pi\mathcal{X}  )} d{\mathcal{X}} \\
 & = \frac{1}{2N( S_{k,l}^0 - S_{k,l}^1  )}\int^{\frac{i-S_{k,l}^1}{N}}_{\frac{i-S_{k,l}^0}{N}} \frac{\sin^2( N\pi\mathcal{X} )}{\sin^2( \pi\mathcal{X} )} d{\mathcal{X}}.
\end{aligned}
\end{equation}
Note that $\mathbb{R} ( [{\bf{\mathring h}}_k]_i )$ and $\mathbb{I} ( [{\bf{\mathring h}}_k]_i )$ are independent. \hfill\rule{6pt}{6pt}

\section*{Appendix B \\ Proof of {\textbf{Proposition 1}}}
The sum-rate of MBMRF precoding can be expressed as
\begin{equation}
\begin{aligned}
\mathbb{E} (R_{\text{MBMRF}}) &\overset{}{=} \mathbb{E} \left\{ \sum^{K}_{k = 1}{\log}_2{ \left( 1+  \frac{ P_{\text{T}} \left( {\bf{\mathring h}}^H_k {\bf{\Xi}}_k {\bf{\Xi}}_k^H {\bf{\mathring h}}_k \right)^2}{ K \sigma ^2 \left\| {\bf{\mathring h}}^H_k {\bf{\Xi}}_k  \right\|^2  }    \right)}   \right\} \\
&\overset{}{=}  \mathbb{E} \left\{ \sum^{K}_{k = 1}{\log}_2{ \left( 1+  \frac{P_{\text{T}} \left( {\bf{\mathring h}}^H_k {\bf{\Xi}}_k {\bf{\Xi}}_k^H {\bf{\mathring h}}_k \right)}{K \sigma ^2}    \right)}   \right\} \\
&\overset{}{=}  \mathbb{E} \left\{ \sum^{K}_{k = 1}{\log}_2{ \left( 1+  \frac{P_{\text{T}}   }{K \sigma ^2 } \sum_{i \in \mathcal{B}_k}  \left[ {\bf{\mathring h}}_k\right]^2_i   \right)}   \right\},
\end{aligned}
\end{equation}
where ${\bf{\Xi}}_k$ is the beam selection matrix for the \emph{k}-th user. In each column of ${\bf{\Xi}}_k$, there is one non-zero element ${1}/{\sqrt{B_k}}$ in the position of selected beams and zeroes in other positions. Then, according to the Jensen's inequality \cite{zi16energy}, we have
\begin{equation}\label{jensen1}
\begin{aligned}
\mathbb{E} (R_{\text{MBMRF}}) & \leq  \hat R_{\text{MBMRF}} \\
    &=  \sum^{K}_{k = 1}{\log}_2{ \left( 1+  \frac{P_{\text{T}}   }{K \sigma ^2 } \sum_{i \in \mathcal{B}_k} \mathbb{E} \left\{ \left[ {\bf{\mathring h}}_k\right]^2_i  \right\} \right) }.\
\end{aligned}
\end{equation}
Recalling \textbf{Lemma 1}, we have
\begin{equation}\label{expec1}
\begin{aligned}
\mathbb{E} \left\{ \left[ {\bf{\mathring h}}_k\right]^2_i  \right\} = \frac{2 N \mu_k \sigma_k^2}{N^k_{\text{cl}}},
\end{aligned}
\end{equation}
which leads to
\begin{equation}
\begin{aligned}
\hat R_{\text{MBMRF}} = \sum^{K}_{k = 1}{\log}_2{ \left( 1+  \frac{ 2 B_k \gamma \sigma^2_k }{  \sigma^2 }   \right) }.
\end{aligned}
\end{equation}
This completes the proof of \textbf{Proposition 1}. \hfill\rule{6pt}{6pt}

\end{document}